\documentclass[lettersize,journal]{IEEEtran}
\usepackage{amsmath,amsfonts,amssymb}
\usepackage{algorithm}
\usepackage{algpseudocode}   
\usepackage{array}
\usepackage[caption=false,font=footnotesize,labelfont=sf,textfont=sf]{subfig}
\usepackage{textcomp}
\usepackage{stfloats}
\usepackage{url}
\usepackage{verbatim}
\usepackage{graphicx}
\usepackage{cite}
\usepackage{bm}
\usepackage{xcolor}
\usepackage{acronym}
\usepackage{orcidlink}
\usepackage{tikz}
\usetikzlibrary{arrows.meta,calc,positioning,decorations.pathreplacing,patterns,backgrounds}
\usepackage{hyperref}
\hypersetup{hidelinks}
\def\BibTeX{{\rm B\kern-.05em{\sc i\kern-.025em b}\kern-.08em T\kern-.1667em\lower.7ex\hbox{E}\kern-.125emX}}

\algrenewcommand\algorithmicrequire{\textbf{Input:}}
\algrenewcommand\algorithmicensure{\textbf{Output:}}

\newtheorem{theorem}{Theorem}
\newtheorem{lemma}{Lemma}
\newtheorem{corollary}{Corollary}
\newtheorem{proposition}{Proposition}

\newcommand{\vect}[1]{\mathbf{#1}}
\newcommand{\gbar}{\bar{\mathbf{g}}}
\newcommand{\Th}{\bm{\Theta}}
\newcommand{\dsin}{\Delta_s}
\newcommand{\Sp}{\mathcal{S}}
\newcommand{\Var}{\mathrm{Var}}

\renewcommand{\Re}{\mathfrak{Re}}
\renewcommand{\Im}{\mathfrak{Im}}

\definecolor{myblue}{RGB}{30,100,180}
\definecolor{myred}{RGB}{190,45,45}
\definecolor{linkblue}{RGB}{32,105,185}
\definecolor{linkred}{RGB}{195,48,48}
\definecolor{panelblue}{RGB}{55,110,175}
\definecolor{trackgray}{RGB}{105,110,120}

\begin{document}

\title{Secure Integrated Sensing and Communications with Movable Beyond-Diagonal RIS}

\author{Chandan~Kumar~Sheemar,~\IEEEmembership{Member,~IEEE,} and Symeon~Chatzinotas,~\IEEEmembership{Fellow,~IEEE}%
\thanks{C. K. Sheemar and S. Chatzinotas are with the Interdisciplinary Centre for Security, Reliability and Trust (SnT), University of Luxembourg, 1855 Luxembourg City, Luxembourg (e-mail: chandankumar.sheemar@uni.lu; symeon.chatzinotas@uni.lu).}}

\maketitle

 \begin{abstract}
 In integrated sensing and communication (ISAC), when the sensed target is itself an eavesdropper, the illumination required for sensing coincides with the leakage that secrecy must suppress. This paper shows how a beyond-diagonal reconfigurable intelligent surface (BD-RIS) with movable elements decouples the two tasks in a monostatic setting, where the echo returns to the base station through the surface. We prove that a lossless reciprocal BD-RIS maps $K$ incident fields onto $K$ prescribed fields if and only if the two sets share the same Gram matrix and their bilinear Gram matrix is symmetric; unlike a diagonal RIS, it can thus steer a two-dimensional incident subspace onto any prescribed subspace. With a single stream, sensing incurs an unavoidable rate loss, whereas a closed-form zero-forcing design with a confidential and a Gaussian sensing stream nulls the data at the target and the sensing at the user, so that sensing costs only an SNR offset. Since all variables follow in closed form for any placement, the joint design reduces to a monotone search over the element positions, without alternating optimization. Simulations show that the dual-stream design loses only $1.1$ bit/s/Hz when the sensing requirement grows by $30$ dB, against $8.3$ bit/s/Hz for a single stream, while movement, interconnection, and the robust derivative-null design add up to $1.9$, $0.9$, and $2.6$ bit/s/Hz, respectively.
\end{abstract} 

\begin{IEEEkeywords}
Movable beyond-diagonal RIS, integrated sensing and communication, physical-layer security.
\end{IEEEkeywords}

\section{Introduction}
\IEEEPARstart{I}{ntegrated} sensing and communication (ISAC) is expected to be a defining feature of sixth-generation (6G) networks: the same waveform, spectrum, and aperture serve users and probe targets~\cite{sheemar2025holographic,sheemar2023full_ris}. If the sensed target is an unauthorized device, e.g., a drone carrying a malicious receiver, the probing signal may reveal the confidential information it carries~\cite{Su2021,Wei2022}. Physical-layer security (PLS)~\cite{Wyner1975} protects against such eavesdroppers by beamforming away from them and radiating artificial noise (AN) toward them~\cite{Goel2008,Khisti2010}, which conflicts with sensing when the eavesdropper is the target. In the seminal treatment of this scenario~\cite{Su2021}, the radar waveform doubles as AN, so that the target is illuminated without decoding the data.

Reconfigurable intelligent surfaces (RISs) extend secrecy and ISAC to blocked geometries~\cite{li2025ris}, and RIS-aided PLS is well studied for conventional, diagonal surfaces~\cite{Cui2019,Guan2020,Yu2020}. Beyond-diagonal RISs (BD-RISs) interconnect their elements through a reconfigurable impedance network and are described, in the lossless reciprocal case, by symmetric unitary scattering matrices~\cite{khan2026survey}; single-, group-, and fully-connected architectures trade circuit complexity for wave control~\cite{Li2023modes,Nerini2024graph}. In parallel, movable and fluid antennas have turned the element positions into a design variable~\cite{iacovelli2026mutual}, and movable-element RISs exploit the same freedom at the surface~\cite{Zhao2026}. In \cite{Xu2026}, the authors combined both freedoms in a movable BD-RIS for the multiuser sum rate.

\subsection{Related Work} 
Secure dual-functional radar-communication designs treat the target as a potential eavesdropper and optimize the beamformers under sensing constraints, using the radar signal or dedicated AN to jam the target~\cite{Su2021,Su2022,Chu2023}. Sensing has been exploited to estimate the eavesdropper~\cite{Su2024}, and robust designs account for imperfect target knowledge~\cite{Ren2023robust}. RIS-aided secure ISAC has been studied with diagonal surfaces, including robust designs for an uncertain target location~\cite{Hua2024secure}, leakage suppression over the angular uncertainty region~\cite{ZhaoH2026}, and transmitting-and-reflecting surfaces~\cite{Wei2024STAR}. Recently, a fixed-element BD-RIS was optimized for secure ISAC with targets as eavesdroppers~\cite{Illi2026}, and a movable diagonal RIS was designed under imperfect sensing estimation~\cite{Zhuang2026}. \emph{These works rely mainly on iterative numerical optimization and do not characterize which pairs of reflected fields a surface can produce simultaneously, which determines whether sensing and secrecy can be decoupled.}

The literature is also evolving considerably towards BD-RIS and movable elements. For single-stream links, the optimal BD-RIS follows in closed form~\cite{Nerini2024closed}. For ISAC, fixed-element BD-RISs improve the throughput~\cite{Liu2024BDISAC}, the Cram\'er--Rao bound~\cite{ZhangX2025CRB}, and the rate--Cram\'er--Rao bound trade-off~\cite{Zhang2025} over diagonal RISs, and BD-RIS-assisted MIMO wiretap channels were studied in~\cite{Xiong2025,ZhangX2025PLS}. In parallel, movable antennas have shown to improve capacity~\cite{Ma2024MIMO}, secrecy~\cite{Hu2024}, and ISAC~\cite{Lyu2025MAISAC}, and have been combined with RISs for secure ISAC~\cite{Ma2025secure}; fluid RISs were proposed for secrecy~\cite{Zhu2025FRIS}. Notably, none of these works study a movable BD-RIS for secure ISAC.

\subsection{Open Problem and Contributions}
For secure ISAC with the target as eavesdropper, the two tasks conflict at the surface: the secrecy rate improves as the correlation $\rho$ between the user and target channels seen by the surface vanishes, whereas illuminating the target with a beam aimed at the user requires $\rho\to1$. With one stream, moving the elements only trades one task against the other, and interconnection cannot help, since one incident field is reflected as one field. The question is what a BD-RIS can do that a diagonal RIS cannot when two fields must be shaped at once, which requires a characterization of the fields a symmetric unitary matrix can realize, and how such null-based designs behave when the target direction, obtained by sensing, is imperfect. To address these gaps, the main contributions of our work are:
 \begin{itemize}
\item Firstly, we characterize the field transformations that a lossless reciprocal BD-RIS can realize. We prove that a symmetric unitary scattering matrix maps $K$ incident fields to $K$ prescribed fields if and only if the two sets share their Gram matrix and their bilinear Gram matrix is symmetric, and the proof is constructive, yielding the scattering matrix in closed form (Theorem~\ref{thm:real}). It is shown that a fully-connected BD-RIS can thereby steer a two-dimensional incident subspace onto any prescribed subspace, whereas a D-RIS cannot and a group-connected BD-RIS must satisfy the conditions group by group.
\item Secondly, we formulate the secure-ISAC problem with monostatic sensing, where the echo returns to the base station through the surface and its return gain is determined by the same frame that shapes the forward fields (Lemma~\ref{lem:ret}). We show that, with a single stream, the sensing requirement costs at least $\log_2(1+\bar\Gamma/2)$ bit/s/Hz for any placement and any surface (Theorem~\ref{thm:one}). To overcome this limitation, we propose a dual-stream design that transmits a confidential stream and a Gaussian sensing stream over a rank-two link, and derive a closed-form zero-forcing (ZF) solution in which the BD-RIS nulls the data at the target and the sensing signal at the user. In particular, it is shown that the sensing requirement then reduces to an SNR offset and that decorrelating the user from the target benefits both tasks.
\item Moreover, we investigate the role of element movement in providing this decorrelation. We derive a double-null placement that attains the interference-free rate in closed form (Proposition~\ref{prop:dn}), lower-bound the gain of movement at high SNR in closed form (Proposition~\ref{prop:gain}), and show that, for a nearby eavesdropper, the gain is governed by the variance of the element positions (Proposition~\ref{prop:small}). In addition, we analyze the robustness of the design to target-angle errors: the same variance renders the nominal null fragile (Proposition~\ref{prop:leak}), and we propose a derivative-null design that reduces the leakage from $O(\delta^2)$ to $O(\delta^4)$ while the sensing stream masks the residual leakage (Proposition~\ref{prop:mask}).
\item Finally, we propose a low-complexity coordinate search for the element positions, in which all remaining variables follow in closed form, and prove that it is monotone and converges in a finite number of sweeps (Proposition~\ref{prop:conv}). We present extensive numerical results under Rician fading to demonstrate the gains of the proposed movable BD-RIS over fixed and diagonal benchmarks\footnote{\emph{Notation:}
$(\cdot)^T$, $(\cdot)^*$, and $(\cdot)^H$ denote the transpose, conjugate, and conjugate transpose, $\|\cdot\|$ the Euclidean norm, and $\mathrm{span}(\cdot)$ the column space. $\vect{P}_{\vect{Z}}^\perp=\vect{I}-\vect{Z}(\vect{Z}^H\vect{Z})^{-1}\vect{Z}^H$ projects onto the orthogonal complement of $\mathrm{span}(\vect{Z})$. $\kappa=2\pi/\lambda$, $[x]^+=\max(x,0)$, and $\Var(\vect{p})=\frac1M\sum_m(p_m-\bar p)^2$ with $\bar p=\frac1M\sum_mp_m$. $\bar\gamma=P/\sigma^2$ and $\bar\Gamma=\Gamma/\sigma^2$ are the reference SNR and the normalized sensing requirement.}. 
\end{itemize}

\emph{Paper Organization:} The remainder of this paper is organized as follows. Section~\ref{sec:model} presents the system model and problem formulation. Section~\ref{sec:real} characterizes the realizable fields of a BD-RIS. Sections~\ref{sec:perfect} and~\ref{sec:robust} derive the secure-ISAC designs with perfect and uncertain target information, respectively. Section~\ref{sec:alg} presents the algorithm, Section~\ref{sec:sim} the numerical results, and Section~\ref{sec:concl} the conclusions.

\section{System Model and Problem Formulation}
\label{sec:model}
\begin{figure}[t]
\centering
\includegraphics[width=\columnwidth]{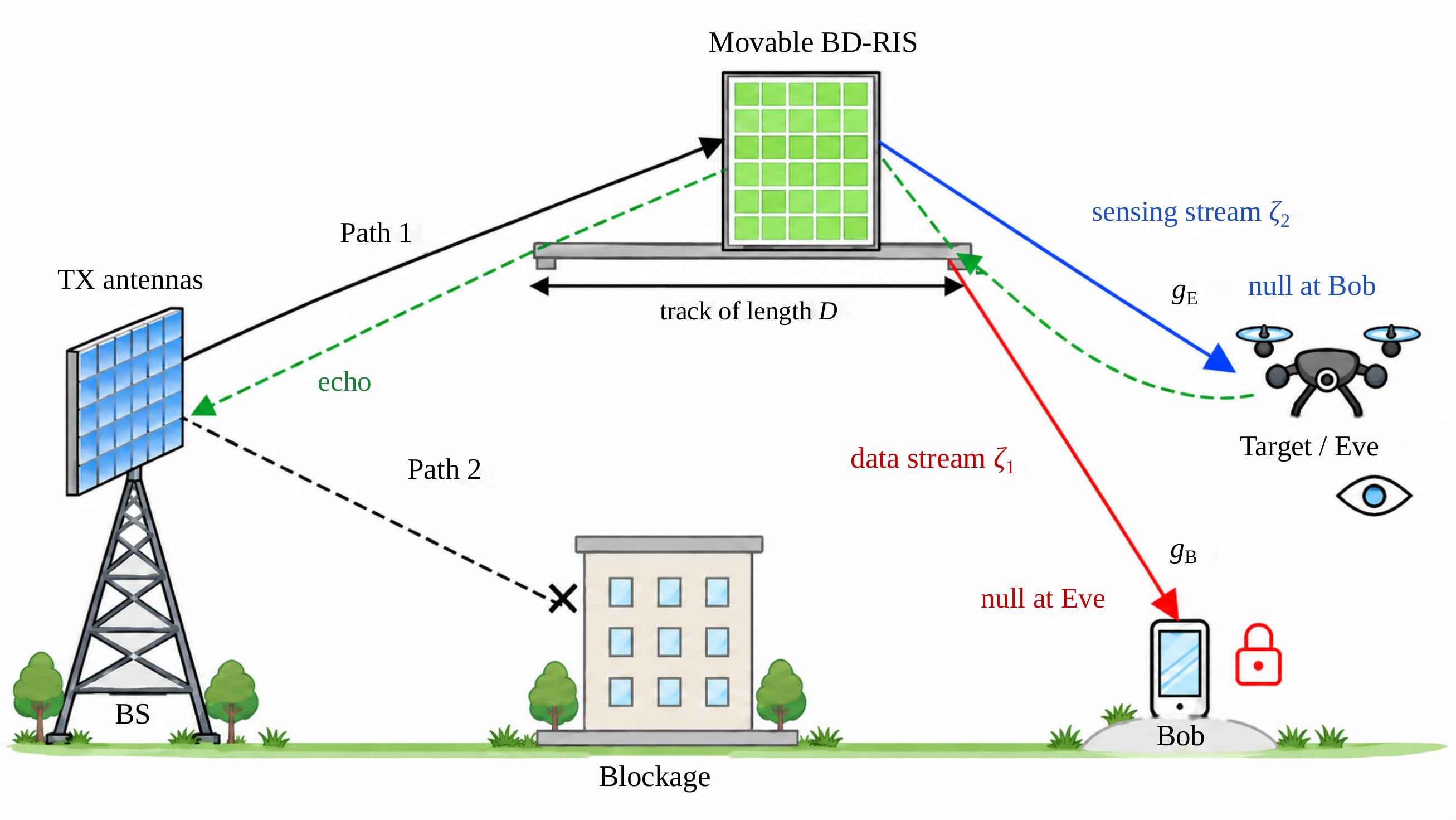}
\caption{Secure ISAC with a movable BD-RIS.}
\label{fig:scenario}
\end{figure}
As illustrated in Fig.~\ref{fig:scenario}, a base station (BS) with an $N$-antenna uniform linear array (ULA) sends a confidential stream to a single-antenna legitimate user Bob (B) and must illuminate a target with a prescribed power. The target carries a single-antenna eavesdropper Eve (E), e.g., an unauthorized device whose position must be tracked. The direct BS--B and BS--E links are blocked, and a BD-RIS with $M$ movable elements assists both functions.

\subsection{Movable BD-RIS}
Element $m$ slides along a linear track of length $D$ at position $p_m$, and the position vector $\vect{p}=[p_1,\dots,p_M]^T$ belongs to
\begin{equation}
\mathcal{X}=\{\vect{p}\in\mathbb{R}^M:\ p_1\geq0,\ p_{m+1}-p_m\geq d_0,\ p_M\leq D\}.
\label{eq:X}
\end{equation}
where $D$ is the length of the track, and $d_0$ is the minimum inter-element spacing.
The impedance network is assumed to follow the elements through flexible feeds, and its $M(M+1)/2$ reactances are assumed lossless. 
The elements are connected to an $M$-port reconfigurable impedance network with scattering matrix $\Th=(\vect{Z}_I+Z_0\vect{I}_M)^{-1}(\vect{Z}_I-Z_0\vect{I}_M)$~\cite{Shen2022}; for a lossless reciprocal network, $\vect{Z}_I=j\vect{X}_I$ with $\vect{X}_I$ real symmetric, so $\Th$ is symmetric and unitary, and in the fully-connected architecture every such matrix is realizable~\cite{Shen2022,Li2023modes}:
\begin{equation}
\mathcal{T}=\{\Th\in\mathbb{C}^{M\times M}:\ \Th=\Th^T,\ \Th^H\Th=\vect{I}_M\}.
\label{eq:T}
\end{equation}
A diagonal RIS (D-RIS) is the subset of diagonal matrices in $\mathcal{T}$, and a group-connected BD-RIS with groups of $M_g$ elements is the subset of block-diagonal matrices with $M_g\times M_g$ symmetric unitary blocks~\cite{Li2023modes}.

\subsection{Channel Model}
We adopt the field-response model~\cite{Zhu2024model}, in which positions enter only through the phase of each planar path. The steering vector of the surface for a path with direction sine $s$ (measured from the broadside) is $\vect{b}(\vect{p};s)$, with $[\vect{b}(\vect{p};s)]_m=e^{j\kappa p_ms}$. The BS--RIS link has two dominant paths, e.g., a line-of-sight (LoS) path and a specular reflection, which arrive at the surface with direction sines $s_{0,1}$ and $s_{0,2}$ and depart from the BS with steering vectors $\vect{a}_1$ and $\vect{a}_2\in\mathbb{C}^N$:
\begin{equation}
\vect{H}(\vect{p})=\sum_{l=1}^{2}\sqrt{\beta_{0,l}}\,\vect{b}(\vect{p};s_{0,l})\,\vect{a}_l^H=\vect{B}(\vect{p})\vect{D}_0^{1/2}\vect{A}^H,
\label{eq:H}
\end{equation}
where $\vect{B}=[\vect{b}(\vect{p};s_{0,1}),\vect{b}(\vect{p};s_{0,2})]$, $\vect{A}=[\vect{a}_1,\vect{a}_2]$, $\vect{D}_0=\mathrm{diag}(\beta_{0,1},\beta_{0,2})$, and $\beta_{0,l}$ are the path gains. The compact singular-value decomposition (SVD) is $\vect{H}=\vect{E}\bm{\Lambda}^{1/2}\vect{W}^H$, with $\vect{E}\in\mathbb{C}^{M\times2}$ and $\vect{W}\in\mathbb{C}^{N\times2}$ having orthonormal columns and $\bm{\Lambda}=\mathrm{diag}(\lambda_1,\lambda_2)$, $\lambda_1\geq\lambda_2\geq0$. The columns of $\vect{E}$ span the \emph{incident subspace} of the surface.

The RIS--$X$ links, $X\in\{\mathrm{B},\mathrm{E}\}$, comprise an LoS path and $L$ scattered paths:
\begin{equation}
\vect{g}_X(\vect{p})=\sqrt{\beta_1}\,\gbar_X(\vect{p}),\quad \gbar_X(\vect{p})=\sum_{l=0}^{L}\alpha_{X,l}\,\vect{b}(\vect{p};s_{X,l}),
\label{eq:g}
\end{equation}
where $\beta_1$ is the large-scale gain, $s_{X,0}$ is the LoS direction sine, $\alpha_{X,0}=\sqrt{K_R/(K_R+1)}$, and $\alpha_{X,l}\sim\mathcal{CN}(0,1/((K_R+1)L))$ for $l\geq1$, with Rician factor $K_R$. Under LoS ($K_R\to\infty$), $\gbar_X=\vect{b}(\vect{p};s_{X,0})$ and $\|\gbar_X\|^2=M$ for every placement. We define the LoS sine differences $\dsin=s_{\mathrm{B},0}-s_{\mathrm{E},0}$ and $\dsin^0=s_{0,1}-s_{0,2}$.

 \subsection{Signal Model and Sensing Metric}
The BS transmits $\vect{x}=\vect{w}_1\varsigma_1+\vect{w}_2\varsigma_2$, where $\varsigma_1$ is the confidential symbol intended for Bob, $\varsigma_2$ is a sensing symbol, both $\mathcal{CN}(0,1)$ and independent, and $\vect{w}_1,\vect{w}_2\in\mathbb{C}^N$ are the transmit precoders of the data and sensing streams, respectively, subject to the total power constraint $\|\vect{w}_1\|^2+\|\vect{w}_2\|^2\leq P$. The sensing sequence is pseudo-random Gaussian, known to the BS for echo processing and unknown to all other nodes~\cite{Su2021}. The incident and reflected fields of stream $k$ are
\begin{equation}
\vect{u}_k=\vect{H}\vect{w}_k,\qquad \vect{v}_k=\Th\vect{u}_k,\qquad k\in\{1,2\},
\label{eq:fields}
\end{equation}
and the signal received by node $X$ is
\begin{equation}
y_X=\vect{g}_X^H\Th\vect{H}\vect{x}+n_X=\vect{g}_X^H\vect{v}_1\varsigma_1+\vect{g}_X^H\vect{v}_2\varsigma_2+n_X,
\label{eq:y}
\end{equation}
with $n_X\sim\mathcal{CN}(0,\sigma^2)$. Sensing is monostatic: the BS operates in full duplex with ideal self-interference suppression, and since its direct link to the target is blocked, the echo returns over the target--RIS--BS path. Modeling the target as a point scatterer with reflection coefficient $\alpha$, the reciprocal target--RIS channel is $\vect{g}_\mathrm{E}^*$, and, since $\Th^T=\Th$, the BS receives
\begin{equation}
\vect{y}_\mathrm{S}=\alpha\vect{H}^T\Th\vect{g}_\mathrm{E}^*\vect{g}_\mathrm{E}^H\Th\vect{H}\vect{x}+\vect{n}_\mathrm{S}=\alpha\vect{h}_\mathrm{E}^*\vect{h}_\mathrm{E}^H\vect{x}+\vect{n}_\mathrm{S},
\label{eq:echo}
\end{equation}
where $\vect{h}_\mathrm{E}=\vect{H}^H\Th^H\vect{g}_\mathrm{E}$ is the effective BS--target channel, $\vect{n}_\mathrm{S}\sim\mathcal{CN}(\vect{0},\sigma^2\vect{I}_N)$, and residual self-interference and clutter are absorbed in $\vect{n}_\mathrm{S}$. By reciprocity, the return channel is the conjugate of the forward one. The BS knows $\vect{x}$, so after receive combining with $\vect{h}_\mathrm{E}^*$ and matched filtering over $T$ snapshots, the echo SNR is~\cite{LiuX2020,Hua2023}
\begin{equation}
\mathrm{SNR}_\mathrm{S}=\frac{\xi\|\vect{h}_\mathrm{E}\|^2P_T}{\sigma^2},\quad P_T=\mathbb{E}|\vect{h}_\mathrm{E}^H\vect{x}|^2=\sum_{k=1}^2|\vect{g}_\mathrm{E}^H\vect{v}_k|^2,
\label{eq:PT}
\end{equation}
where $\xi=T\mathbb{E}|\alpha|^2$ collects the radar cross section and the integration gain. The echo SNR is the product of the illumination $P_T$ on the forward path and the return gain $\|\vect{h}_\mathrm{E}\|^2$ on the backward path, and both depend on $\Th$ and $\vect{p}$. We require $\mathrm{SNR}_\mathrm{S}\geq\Gamma_\mathrm{S}$, i.e.,
\begin{equation}
\|\vect{h}_\mathrm{E}\|^2P_T\geq\Gamma h_\mathrm{ref},\qquad h_\mathrm{ref}=\beta_0\beta_1NM^2,
\label{eq:sens}
\end{equation}
with $\Gamma=\Gamma_\mathrm{S}\sigma^2/(\xi h_\mathrm{ref})$. Here $h_\mathrm{ref}$ is the return gain of an isotropic incident subspace under LoS (Lemma~\ref{lem:ret}), so $\Gamma$ is the illumination needed in that reference case.

\subsection{Secrecy Metric and Threat Model}
Eve is a passive eavesdropper that knows the codebook of the confidential stream and $\vect{g}_\mathrm{E}^H\vect{v}_1$, but not the realization of the sensing sequence. Since $\varsigma_1$ and $\varsigma_2$ are independent Gaussian, $I(\varsigma_1;y_X)=\log_2(1+\gamma_X)$, with
\begin{equation}
\gamma_X=\frac{|\vect{g}_X^H\vect{v}_1|^2}{\sigma^2+|\vect{g}_X^H\vect{v}_2|^2},\qquad X\in\{\mathrm{B},\mathrm{E}\},
\label{eq:sinr}
\end{equation}
so the sensing stream acts as AN at both receivers. By the Gaussian wiretap channel~\cite{Wyner1975,Csiszar1978}, the achievable secrecy rate is
\begin{equation}
R_s=\big[\log_2(1+\gamma_\mathrm{B})-\log_2(1+\gamma_\mathrm{E})\big]^+.
\label{eq:Rs}
\end{equation}
In Sections~\ref{sec:model}--\ref{sec:perfect}, all channels are assumed known at the BS. For the target, this models the tracking regime of ISAC, in which the BS has localized the target through previous echoes~\cite{Su2024}. In Section~\ref{sec:robust}, the target direction is known only up to an error $\delta$.

\subsection{Problem Formulation}
The joint design problem of the precoders, the scattering matrix, and the element positions to maximize the secrect rate is
\begin{align}
(\mathrm{P1}):\ \max_{\vect{w}_1,\vect{w}_2,\Th,\vect{p}}\ \ & R_s\nonumber\\
\mathrm{s.t.}\ \ & \|\vect{h}_\mathrm{E}\|^2P_T\geq\Gamma h_\mathrm{ref},\nonumber\\ & \|\vect{w}_1\|^2+\|\vect{w}_2\|^2\leq P,\nonumber\\
& \Th\in\mathcal{T},\quad \vect{p}\in\mathcal{X}.
\label{eq:P1}
\end{align}
Problem (P1) is nonconvex in all variables, so we first ask which fields $(\vect{v}_1,\vect{v}_2)$ the surface can realize, and then design the fields.

Let $\rho=|\vect{g}_\mathrm{B}^H\vect{g}_\mathrm{E}|/(\|\vect{g}_\mathrm{B}\|\|\vect{g}_\mathrm{E}\|)$ be the Bob--Eve correlation at the surface and $\Sp=\mathrm{span}\{\vect{g}_\mathrm{B},\vect{g}_\mathrm{E}\}$. Since all metrics depend on $\vect{v}_k$ only through $\vect{g}_X^H\vect{v}_k$, field energy outside $\Sp$ is wasted. Under LoS, $\rho(\vect{p})=|A(\vect{p};\dsin)|$ with the array factor
\begin{equation}
A(\vect{p};s)=\frac1M\sum_{m=1}^Me^{j\kappa sp_m},
\label{eq:AF}
\end{equation}
so the placement controls $\rho$ directly. For the compact array, $|A(\vect{p};s)|=|\sin(M\pi d_0s/\lambda)/(M\sin(\pi d_0s/\lambda))|$.

\section{Field Realizability of a BD-RIS}
\label{sec:real}
This section characterizes the mappings between incident and reflected fields that a BD-RIS can realize. The results hold for any channel model and are of independent interest for multi-field BD-RIS design.

\subsection{Main Result}
\begin{theorem}[Realizability]
\label{thm:real}
Let $\vect{U},\vect{V}\in\mathbb{C}^{M\times K}$. There exists $\Th\in\mathcal{T}$ with $\Th\vect{U}=\vect{V}$ if and only if
\begin{equation}
\mathrm{(i)}\ \ \vect{U}^H\vect{U}=\vect{V}^H\vect{V},\qquad \mathrm{(ii)}\ \ \vect{U}^T\vect{V}=\vect{V}^T\vect{U}.
\label{eq:cond}
\end{equation}
Moreover, $\Th=\vect{Q}\vect{Q}^T$, where $\vect{Q}$ is any unitary matrix with $\vect{Q}[\widetilde{\vect{U}},\widetilde{\vect{U}}^*]=[\vect{V},\vect{U}^*]$ and $\widetilde{\vect{U}}$ is constructed in closed form in Appendix~\ref{app:real}.
\end{theorem}
\begin{IEEEproof}
Necessity: (i) follows from $\Th^H\Th=\vect{I}_M$, and $\vect{U}^T\vect{V}=\vect{U}^T\Th\vect{U}$ is symmetric because $\Th$ is. The constructive proof of sufficiency, based on the Takagi factorization and a real embedding, is given in Appendix~\ref{app:real}.
\end{IEEEproof}

Condition (i) is energy conservation, and (ii) is the signature of reciprocity: the bilinear cross-products $\vect{u}_i^T\vect{v}_j$ must be symmetric.

\subsection{Consequences}
\begin{corollary}[One field]
\label{cor:one}
For $K=1$, (ii) is void, so $\Th\vect{u}=\vect{v}$ is realizable if and only if $\|\vect{v}\|=\|\vect{u}\|$.
\end{corollary}
This recovers the fact that a fully-connected BD-RIS steers one field anywhere~\cite{Nerini2024closed}. For $K=2$, (ii) is the scalar equation $\vect{u}_1^T\vect{v}_2=\vect{u}_2^T\vect{v}_1$, which constrains the simultaneous steering of two fields.

\begin{corollary}[Subspace steering]
\label{cor:sub}
Let $\vect{E},\vect{Q}_0\in\mathbb{C}^{M\times2}$ have orthonormal columns and let $\vect{E}^T\vect{Q}_0=\bm{\Phi}\bm{\Sigma}\vect{R}^H$ be an SVD with distinct singular values. Then $\Th\vect{E}=\vect{F}$ with $\mathrm{span}(\vect{F})=\mathrm{span}(\vect{Q}_0)$ is realizable by some $\Th\in\mathcal{T}$ if and only if
\begin{equation}
\vect{F}=\vect{Q}_0\vect{R}\,\mathrm{diag}(e^{j\theta_1},e^{j\theta_2})\,\bm{\Phi}^T,\qquad \theta_1,\theta_2\in[0,2\pi).
\label{eq:frame}
\end{equation}
In particular, the incident subspace of $\vect{H}$ can be mapped onto $\Sp$, or onto any other two-dimensional subspace.
\end{corollary}
\begin{IEEEproof}
By (i), $\vect{F}^H\vect{F}=\vect{E}^H\vect{E}=\vect{I}_2$, so $\vect{F}=\vect{Q}_0\bm{\Psi}$ with $\bm{\Psi}$ a $2\times2$ unitary matrix, which we write as $\bm{\Psi}=\vect{R}\bm{\Omega}\bm{\Phi}^T$ with $\bm{\Omega}$ unitary. Condition (ii) requires $\vect{E}^T\vect{F}=\bm{\Phi}\bm{\Sigma}\bm{\Omega}\bm{\Phi}^T$ to be symmetric, i.e., $\bm{\Sigma}\bm{\Omega}=\bm{\Omega}^T\bm{\Sigma}$. Comparing the off-diagonal entries gives $\sigma_1\omega_{12}=\sigma_2\omega_{21}$. Since $|\omega_{12}|=|\omega_{21}|$ for every $2\times2$ unitary matrix, $\sigma_1\neq\sigma_2$ forces $\omega_{12}=\omega_{21}=0$, so $\bm{\Omega}$ is diagonal with unit-modulus entries. Conversely, every such $\bm{\Omega}$ satisfies (i)--(ii), and Theorem~\ref{thm:real} applies.
\end{IEEEproof}

Only $\phi=\theta_2-\theta_1$ matters for the design, and an incident field $\vect{u}=\vect{E}\vect{t}$ is reflected as $\vect{v}=\vect{F}\vect{t}$. If $\sigma_1=\sigma_2$, any symmetric unitary $\bm{\Omega}$ is admissible and the family is larger; the designs below use only~\eqref{eq:frame}, which remains realizable in that case.

\begin{proposition}[Limits of a D-RIS]
\label{prop:dris}
A D-RIS realizes $\Th\vect{U}=\vect{V}$ only if $|V_{mk}|=|U_{mk}|$ and $V_{m1}U_{m2}=V_{m2}U_{m1}$ for all $m$ and $k$. For two generic incident fields and two generic target fields in $\Sp$, these $3M$ real conditions cannot be met by the $M$ phases of the D-RIS.
\end{proposition}
\begin{IEEEproof}
For $\Th=\mathrm{diag}(e^{j\vartheta_1},\dots,e^{j\vartheta_M})$, $V_{mk}=e^{j\vartheta_m}U_{mk}$, which gives both conditions. The modulus conditions are $2M$ real equations and the ratio condition $M$ further real equations (its modulus part is implied), against $M$ phases.
\end{IEEEproof}

\begin{corollary}[Group-connected BD-RIS]
\label{cor:group}
Partition the elements into groups $\mathcal{G}_1,\dots,\mathcal{G}_G$ of size $M_g$, and let $\vect{U}_g,\vect{V}_g$ denote the rows of $\vect{U},\vect{V}$ indexed by $\mathcal{G}_g$. A group-connected BD-RIS realizes $\Th\vect{U}=\vect{V}$ if and only if $\vect{U}_g^H\vect{U}_g=\vect{V}_g^H\vect{V}_g$ and $\vect{U}_g^T\vect{V}_g$ is symmetric for every $g$.
\end{corollary}
\begin{IEEEproof}
A block-diagonal $\Th$ acts independently on each group, so Theorem~\ref{thm:real} applies to each block.
\end{IEEEproof}
Each group must preserve the Gram and bilinear Gram matrices of its own rows, which generically excludes subspace steering when $G\geq2$; for $M_g=1$ the conditions reduce to Proposition~\ref{prop:dris}, and the fully-connected surface ($G=1$) imposes the weakest conditions.

\subsection{Incident Fields and the Rank of the BS--RIS Link}
The BS controls only the incident fields, through~\eqref{eq:fields}. The following lemma characterizes them.
\begin{lemma}[Incident fields]
\label{lem:inc}
Let $\lambda_2>0$. The incident field of any stream is $\vect{u}=\vect{E}\vect{t}$ with $\vect{t}\in\mathbb{C}^2$, and the minimum transmit power producing it is
\begin{equation}
\min_{\vect{w}:\vect{H}\vect{w}=\vect{E}\vect{t}}\|\vect{w}\|^2=\vect{t}^H\bm{\Lambda}^{-1}\vect{t}\in\Big[\frac{\|\vect{t}\|^2}{\lambda_1},\frac{\|\vect{t}\|^2}{\lambda_2}\Big],
\label{eq:inc}
\end{equation}
attained by $\vect{w}=\vect{W}\bm{\Lambda}^{-1/2}\vect{t}$. If the paths are resolvable at the BS ($\vect{A}^H\vect{A}=N\vect{I}_2$), then
\begin{equation}
\lambda_{1,2}=\frac{NM}{2}\Big[\beta_{0,1}+\beta_{0,2}\pm\sqrt{(\beta_{0,1}-\beta_{0,2})^2+4\beta_{0,1}\beta_{0,2}\rho_0^2}\Big],
\label{eq:lam}
\end{equation}
where $\rho_0=|A(\vect{p};\dsin^0)|$. For equal gains $\beta_{0,l}=\beta_0$, $\lambda_{1,2}=\beta_0NM(1\pm\rho_0)$.
\end{lemma}
\begin{IEEEproof}
 The proof is provided in Appendix~\ref{app:inc}
\end{IEEEproof}
The rank of the BS--RIS link is thus a resource: with a single path, all incident and hence all reflected fields are parallel for every $\Th$, and data and sensing cannot be separated. A weak second path gives $\lambda_2\approx\beta_{0,2}NM(1-\rho_0^2)$ and makes that mode expensive; movement acts through $\rho_0$, equalizing the modes ($\rho_0\to0$) or concentrating the power in the strongest ($\rho_0\to1$). The same modes govern the echo.
\begin{lemma}[Return gain]
\label{lem:ret}
If $\Th\vect{E}=\vect{F}=[\vect{f}_1,\vect{f}_2]$, then $\|\vect{h}_\mathrm{E}\|^2=\sum_{i=1}^2\lambda_i|\vect{g}_\mathrm{E}^H\vect{f}_i|^2\leq\lambda_1\|\vect{g}_\mathrm{E}\|^2$ for every $\Th\in\mathcal{T}$. If $\vect{g}_\mathrm{E}\in\mathrm{span}(\vect{F})$, then $\|\vect{h}_\mathrm{E}\|^2\geq\lambda_2\|\vect{g}_\mathrm{E}\|^2$, and for an isotropic incident subspace under LoS, $\|\vect{h}_\mathrm{E}\|^2=\lambda_\mathrm{iso}\beta_1M$.
\end{lemma}
\begin{IEEEproof}
$\vect{h}_\mathrm{E}=\vect{W}\bm{\Lambda}^{1/2}\vect{E}^H\Th^H\vect{g}_\mathrm{E}=\vect{W}\bm{\Lambda}^{1/2}\vect{F}^H\vect{g}_\mathrm{E}$, and $\vect{F}$ has orthonormal columns because $\Th$ is unitary.
\end{IEEEproof}
Thus the frame that shapes the forward fields also sets the return gain, a coupling that the design below must respect.

\section{Secure ISAC with Perfect Target Information}
\label{sec:perfect}
This section applies the realizability results of Section~\ref{sec:real} to problem (P1) when all channels, including the target's, are known. We first show that a single stream faces an intrinsic conflict between secrecy and sensing, then derive a closed-form dual-stream design that decouples the two tasks, and finally characterize the element placements that make this design interference-free.
\subsection{Single Stream: An Intrinsic Conflict}
We first consider $\vect{w}_2=\vect{0}$, in which case the confidential stream must also illuminate the target. We send it through the strongest incident mode $\vect{e}_1$ (the first column of $\vect{E}$), which provides the largest field energy $\mathcal{E}\triangleq\lambda_1P$ (Lemma~\ref{lem:inc}). By Corollary~\ref{cor:one}, every $\vect{v}\in\Sp$ with $\|\vect{v}\|^2=\mathcal{E}$ is realizable, and the second mode $\vect{e}_2$ is free to be mapped to any unit $\vect{f}_2$ allowed by Theorem~\ref{thm:real}, which we use to maximize the return gain. Let $a=\|\vect{g}_\mathrm{B}\|^2/\sigma^2$, $b=\|\vect{g}_\mathrm{E}\|^2/\sigma^2$, and $\hat{\vect{g}}_X=\vect{g}_X/\|\vect{g}_X\|$.
\begin{theorem}[Single stream through the strongest mode]
\label{thm:one}
Let $\vect{v}(\ell)=\sqrt{\ell}\,e^{j\angle r}\hat{\vect{g}}_\mathrm{E}+\sqrt{\mathcal{E}-\ell}\,\vect{e}_\perp$, where $\hat{\vect{g}}_\mathrm{B}=r\hat{\vect{g}}_\mathrm{E}+\sqrt{1-\rho^2}\,\vect{e}_\perp$ with $\vect{e}_\perp\perp\hat{\vect{g}}_\mathrm{E}$, and let $\hat{\vect{v}}=\vect{v}/\sqrt{\mathcal{E}}$, $\vect{a}_\perp=\vect{P}^\perp_{\hat{\vect{v}}}\vect{e}_1^*$, and $c=\vect{e}_2^T\hat{\vect{v}}$. The optimum of (P1) with one stream through $\vect{e}_1$, for given positions, is
\begin{equation}
R_s^{(1)}=\max_{\ell\in\mathcal{L}}\log_2\frac{1+a\big(\rho\sqrt{\ell}+\sqrt{(1-\rho^2)(\mathcal{E}-\ell)}\big)^2}{1+b\ell},
\label{eq:one}
\end{equation}
where $\mathcal{L}=\{\ell\in[0,\mathcal{E}]:\ h(\ell)\,b\sigma^2\ell\geq\Gamma h_\mathrm{ref}\}$, $h(\ell)=\lambda_1b\sigma^2\ell/\mathcal{E}+\lambda_2\varpi(\ell)^2$, and
\begin{equation}
\varpi=\frac{|c||\vect{a}_\perp^H\vect{g}_\mathrm{E}|}{\|\vect{a}_\perp\|^2}+\|\vect{P}^\perp_{[\hat{\vect{v}},\vect{a}_\perp]}\vect{g}_\mathrm{E}\|\sqrt{1-\frac{|c|^2}{\|\vect{a}_\perp\|^2}}.
\label{eq:varpi}
\end{equation}
For every placement and every surface, $R_s^{(1)}\leq\log_2(1+a\mathcal{E})-\log_2\big(1+\bar\Gamma h_\mathrm{ref}/(\lambda_1\|\vect{g}_\mathrm{E}\|^2)\big)$; under LoS, the second term is at least $\log_2(1+\bar\Gamma\beta_0NM/\lambda_1)\geq\log_2(1+\bar\Gamma/2)$.
\end{theorem}
\begin{IEEEproof}
Write $\vect{v}=c_\mathrm{E}\hat{\vect{g}}_\mathrm{E}+c_\perp\vect{e}_\perp$ with leakage $\ell=|c_\mathrm{E}|^2$, so $\gamma_\mathrm{E}=b\ell$ and $P_T=b\sigma^2\ell$. Bob's gain $|\hat{\vect{g}}_\mathrm{B}^H\vect{v}|\leq\rho\sqrt{\ell}+\sqrt{1-\rho^2}|c_\perp|$ is attained by phase alignment and maximized by $|c_\perp|^2=\mathcal{E}-\ell$, which gives $\vect{v}(\ell)$ and the numerator of~\eqref{eq:one}. By Theorem~\ref{thm:real}, $\Th\vect{e}_1=\hat{\vect{v}}$ and $\Th\vect{e}_2=\vect{f}_2$ are jointly realizable if and only if $\vect{f}_2\perp\hat{\vect{v}}$, $\|\vect{f}_2\|=1$, and $\vect{e}_1^T\vect{f}_2=c$. Writing $\vect{f}_2=c\vect{a}_\perp/\|\vect{a}_\perp\|^2+\vect{f}_\perp$ with $\vect{f}_\perp\perp\{\hat{\vect{v}},\vect{a}_\perp\}$ and $\|\vect{f}_\perp\|^2=1-|c|^2/\|\vect{a}_\perp\|^2$, the triangle and Cauchy--Schwarz inequalities give $|\vect{g}_\mathrm{E}^H\vect{f}_2|\leq\varpi$, with equality for $\vect{f}_\perp$ aligned in phase with $\vect{P}^\perp_{[\hat{\vect{v}},\vect{a}_\perp]}\vect{g}_\mathrm{E}$. Lemma~\ref{lem:ret} then gives $\|\vect{h}_\mathrm{E}\|^2=h(\ell)$, and~\eqref{eq:sens} defines $\mathcal{L}$. For the bound, $\gamma_\mathrm{B}\leq a\mathcal{E}$ and, by~\eqref{eq:sens} and Lemma~\ref{lem:ret}, $\gamma_\mathrm{E}=P_T/\sigma^2\geq\bar\Gamma h_\mathrm{ref}/\|\vect{h}_\mathrm{E}\|^2\geq\bar\Gamma h_\mathrm{ref}/(\lambda_1\|\vect{g}_\mathrm{E}\|^2)$ for any design; under LoS, $\|\vect{g}_\mathrm{E}\|^2=\beta_1M$ and $\lambda_1\leq2\beta_0NM$ by~\eqref{eq:lam}.
\end{IEEEproof}

Theorem~\ref{thm:one} fixes the incident mode and is therefore a design, not the single-stream optimum of (P1); its bound, however, holds for every single-stream design: the illuminating energy is information-bearing, so Eve's SNR is at least the illumination that the echo needs, and the secrecy rate lies at least $\log_2(1+\bar\Gamma\beta_0NM/\lambda_1)$ bit/s/Hz below Bob's rate bound $\log_2(1+a\mathcal{E})$, for any placement and architecture. Movement can only mitigate this loss, by making the natural leakage of the beam aimed at Bob match the required illumination and by increasing $\lambda_1$ ($\rho_0\to1$), which strengthens the forward and the return path alike.

\subsection{Two Streams: Decorrelation Serves Both Tasks}
With two streams, the incident fields span the two-dimensional incident subspace, and Corollary~\ref{cor:sub} allows the BD-RIS to steer it onto $\Sp$. We use the zero-forcing (ZF) field directions
\begin{equation}
\vect{d}_1=\frac{\vect{P}_{\vect{g}_\mathrm{E}}^\perp\vect{g}_\mathrm{B}}{\|\vect{P}_{\vect{g}_\mathrm{E}}^\perp\vect{g}_\mathrm{B}\|},\qquad \vect{d}_2=\frac{\vect{P}_{\vect{g}_\mathrm{B}}^\perp\vect{g}_\mathrm{E}}{\|\vect{P}_{\vect{g}_\mathrm{B}}^\perp\vect{g}_\mathrm{E}\|},
\label{eq:zf}
\end{equation}
for the data stream (no leakage to Eve) and the sensing stream (no interference to Bob), respectively, and the fields $\vect{v}_k=\sqrt{\mathcal{E}_k}\,\vect{d}_k$. Both directions lie in $\Sp$, and $\|\vect{P}_{\vect{g}_\mathrm{E}}^\perp\vect{g}_\mathrm{B}\|^2=\|\vect{g}_\mathrm{B}\|^2(1-\rho^2)$.

\begin{theorem}[ZF two-stream design]
\label{thm:two}
Let $\vect{Q}_0$ be an orthonormal basis of $\Sp$ and $\vect{F}(\phi)=\vect{Q}_0\vect{R}\,\mathrm{diag}(1,e^{j\phi})\bm{\Phi}^T$ the realizable frame of Corollary~\ref{cor:sub}. The ZF fields with energies $\mathcal{E}_1,\mathcal{E}_2$ are realizable with transmit power $\pi_1(\phi)\mathcal{E}_1+\pi_2(\phi)\mathcal{E}_2$, where
\begin{equation}
\pi_k(\phi)=\vect{d}_k^H\vect{F}(\phi)\bm{\Lambda}^{-1}\vect{F}(\phi)^H\vect{d}_k\in\Big[\frac1{\lambda_1},\frac1{\lambda_2}\Big]
\label{eq:pik}
\end{equation}
is the incident cost of direction $\vect{d}_k$, and $\|\vect{h}_\mathrm{E}(\phi)\|^2=\sum_i\lambda_i|\vect{g}_\mathrm{E}^H\vect{f}_i(\phi)|^2\in[\lambda_2,\lambda_1]\|\vect{g}_\mathrm{E}\|^2$ is the return gain. With $\kappa(\phi)=h_\mathrm{ref}/\|\vect{h}_\mathrm{E}(\phi)\|^2$, the secrecy rate is
\begin{equation}
R_s^{(2)}=\max_{\phi}\log_2\!\Big(1+\frac{\|\vect{g}_\mathrm{B}\|^2(1-\rho^2)P}{\sigma^2\pi_1(\phi)}-\frac{\pi_2\kappa\|\vect{g}_\mathrm{B}\|^2}{\pi_1\|\vect{g}_\mathrm{E}\|^2}\bar\Gamma\Big),
\label{eq:two}
\end{equation}
where $\pi_k$ and $\kappa$ depend on $\phi$. It is feasible if and only if $P\geq\min_\phi\pi_2(\phi)\kappa(\phi)\Gamma/(\|\vect{g}_\mathrm{E}\|^2(1-\rho^2))$, and $\gamma_\mathrm{E}=0$. For an isotropic incident subspace ($\lambda_1=\lambda_2=\lambda_\mathrm{iso}$) under LoS,
\begin{equation}
R_s^{(2)}=\log_2\!\Big(1+\lambda_\mathrm{iso}\beta_1M(1-\rho^2)\bar\gamma-\frac{\beta_0NM}{\lambda_\mathrm{iso}}\bar\Gamma\Big).
\label{eq:twoiso}
\end{equation}
\end{theorem}
\begin{IEEEproof}
By Corollary~\ref{cor:sub}, $\Th\vect{E}=\vect{F}(\phi)$ is realizable, so the incident field $\vect{u}_k=\vect{E}\vect{t}_k$ is reflected as $\vect{F}\vect{t}_k$, and $\vect{v}_k=\vect{F}\vect{t}_k$ holds for $\vect{t}_k=\vect{F}^H\vect{v}_k$ because $\vect{v}_k\in\mathrm{span}(\vect{F})=\Sp$. By Lemma~\ref{lem:inc}, stream $k$ requires the power $\vect{t}_k^H\bm{\Lambda}^{-1}\vect{t}_k=\mathcal{E}_k\pi_k(\phi)$, and $\pi_k\in[1/\lambda_1,1/\lambda_2]$ because $\|\vect{F}^H\vect{d}_k\|=1$. Since $\vect{g}_\mathrm{E}^H\vect{d}_1=0$ and $\vect{g}_\mathrm{B}^H\vect{d}_2=0$, we have $\gamma_\mathrm{E}=0$, $\gamma_\mathrm{B}=\|\vect{g}_\mathrm{B}\|^2(1-\rho^2)\mathcal{E}_1/\sigma^2$, and $P_T=|\vect{g}_\mathrm{E}^H\vect{d}_2|^2\mathcal{E}_2=\|\vect{g}_\mathrm{E}\|^2(1-\rho^2)\mathcal{E}_2$. Since $\Sp=\mathrm{span}(\vect{F})$, Lemma~\ref{lem:ret} gives the return gain and its range. The rate increases in $\mathcal{E}_1$, so the sensing constraint~\eqref{eq:sens} is active, $\mathcal{E}_2=\kappa\Gamma/(\|\vect{g}_\mathrm{E}\|^2(1-\rho^2))$, and the power constraint gives $\mathcal{E}_1=(P-\pi_2\mathcal{E}_2)/\pi_1$. Substituting and maximizing over $\phi$ yields~\eqref{eq:two} and the feasibility condition. For $\lambda_1=\lambda_2=\lambda_\mathrm{iso}$, $\pi_k=1/\lambda_\mathrm{iso}$, $\|\vect{g}_X\|^2=\beta_1M$ under LoS, and $\kappa=\beta_0NM/\lambda_\mathrm{iso}$ by Lemma~\ref{lem:ret}, which gives~\eqref{eq:twoiso}.
\end{IEEEproof}

The dependence on $\phi$ has a simple structure, which we prove in Appendix~\ref{app:phase}.
\begin{lemma}[Phase structure]
\label{lem:phase}
For each $k$, $\pi_k(\phi)=c_{k,0}+2|c_{k,1}|\cos(\phi-\psi_k)$ for constants $c_{k,0}>0$, $c_{k,1}\in\mathbb{C}$, and $\psi_k=\arg c_{k,1}$, with $c_{k,0}\pm2|c_{k,1}|\in[1/\lambda_1,1/\lambda_2]$. The return gain $\|\vect{h}_\mathrm{E}(\phi)\|^2$ is a sinusoid of the same form.
\end{lemma}
Hence the objective of~\eqref{eq:two} is a smooth $2\pi$-periodic function of $\phi$ composed of three sinusoids, and a grid of $N_\phi$ points finds its maximum with a rate error of $O(N_\phi^{-2})$. The phase $\phi$ trades the data cost $\pi_1$ against the sensing cost $\pi_2\kappa$, i.e., against how well the frame returns the echo.

Theorem~\ref{thm:two} has three consequences. First, Eve receives the power $\Gamma$ but no information, since the illumination is carried by the Gaussian sensing stream, which also masks any residual data leakage (Section~\ref{sec:robust}). Second, sensing enters~\eqref{eq:twoiso} as a correlation-independent offset: the sensing energy $\kappa\Gamma/(\|\vect{g}_\mathrm{E}\|^2(1-\rho^2))$ grows with $\rho$ exactly as the data gain $\|\vect{g}_\mathrm{B}\|^2(1-\rho^2)$ shrinks. Third, $R_s^{(2)}$ in~\eqref{eq:twoiso} decreases in $\rho$, so, unlike in Theorem~\ref{thm:one}, the placement should minimize $\rho$: decorrelation serves secrecy and sensing at once. The single-stream design thus suffers a multiplicative SNR loss of at least $(1+\bar\Gamma\beta_0NM/\lambda_1)^{-1}$, the dual-stream design only an additive loss and the factor $1-\rho^2$.

The ZF fields are one feasible pair, so~\eqref{eq:two} is an achievable rate and a lower bound on the two-stream optimum of (P1), which may let the data leak toward Eve when the sensing stream masks it; Section~\ref{sec:sim} quantifies the gap numerically.

\subsection{When Do Two Streams Win?}
\begin{corollary}[Two streams versus one]
\label{cor:win}
Under LoS with resolvable equal-gain paths and $D\geq D_\mathrm{dn}$, let both designs use their optimal placements. Then $\liminf_{\bar\gamma\to\infty}\big(R_s^{(2)}-R_s^{(1)}\big)\geq\log_2\big((1+\bar\Gamma/2)/2\big)$, so two streams do at least as well as one at high SNR whenever $\bar\Gamma\geq2$ ($3$\,dB).
\end{corollary}
\begin{IEEEproof}
By Theorem~\ref{thm:one}, $R_s^{(1)}\leq\log_2(1+\lambda_1\beta_1M\bar\gamma)-\log_2(1+\bar\Gamma\beta_0NM/\lambda_1)$ for every placement. At high SNR this bound increases in $\lambda_1$, and $\lambda_1\leq2\beta_0NM$ by~\eqref{eq:lam}, with equality when the incident paths add coherently ($\rho_0=1$). The double-null placement of Proposition~\ref{prop:dn} below is feasible for $D\geq D_\mathrm{dn}$, so $R_s^{(2)}\geq\log_2(1+\beta_0\beta_1NM^2\bar\gamma-\bar\Gamma)$ by~\eqref{eq:twoiso}. Both bounds grow as $\log_2\bar\gamma$, and their difference tends to $\log_2\big((1+\bar\Gamma/2)/2\big)$.
\end{IEEEproof}
At low SNR the comparison can reverse: the dual-stream design must illuminate the target through the direction orthogonal to Bob, at the cost $1/(1-\rho^2)$, whereas the single-stream beam illuminates the target as a by-product and uses the strongest mode on both paths, so it meets the sensing requirement with less power.

\subsection{Role of Placement}
Under LoS, $\rho=|A(\vect{p};\dsin)|$ and $\rho_0=|A(\vect{p};\dsin^0)|$ are array factors of the placement, and Theorem~\ref{thm:two} favors $\rho=0$ and, for isotropy, $\rho_0=0$.
\begin{proposition}[Double-null placement]
\label{prop:dn}
Let $M=4M_b$, let $\Delta_0$ be the smallest odd multiple of $\lambda/(2|\dsin^0|)$ not smaller than $M_bd_0$, and let $\Delta_1$ be the smallest odd multiple of $\lambda/(2|\dsin|)$ not smaller than $\Delta_0+M_bd_0$. Placing four $d_0$-spaced blocks of $M_b$ elements at offsets $\{0,\Delta_0,\Delta_1,\Delta_0+\Delta_1\}$ is feasible if $D\geq D_\mathrm{dn}\triangleq\Delta_0+\Delta_1+(M_b-1)d_0$ and gives $\rho=\rho_0=0$. With resolvable equal-gain paths, $\lambda_\mathrm{iso}=\beta_0NM$ and
\begin{equation}
R_s^{(2)}=\log_2\!\big(1+\beta_0\beta_1NM^2\bar\gamma-\bar\Gamma\big).
\label{eq:dn}
\end{equation}
Conversely, if $|\dsin|D<\lambda/2$, then $\rho\geq\cos(\pi|\dsin|D/\lambda)>0$ for every $\vect{p}\in\mathcal{X}$, and likewise for $\rho_0$ with $\dsin^0$.
\end{proposition}
\begin{IEEEproof}
See Appendix~\ref{app:dn}.
\end{IEEEproof}
Equation~\eqref{eq:dn} is the interference-free benchmark, with full array gain $M^2$ and Eve nulled: interconnection makes the dual-stream design possible and movement makes it interference-free. Placements with $\rho_0>0$ can exceed it by sending the data through the stronger incident mode, $\lambda_1>\beta_0NM$. Since $\rho$ and $\rho_0$ depend only on the LoS angles, the positions can be updated on the slow angular time scale, while the fields, precoders, and $\Th$ follow the fast fading in closed form.

The next result quantifies the gain of movement at high SNR.
\begin{proposition}[Gain of movement]
\label{prop:gain}
Under LoS with resolvable equal-gain paths, let $\rho_\mathrm{c}$ and $\rho_{0,\mathrm{c}}$ be the correlations of the compact array. If $D\geq D_\mathrm{dn}$, then as $\bar\gamma\to\infty$ the gain of the movable BD-RIS at its optimal placement $\vect{p}^\star$ over the fixed BD-RIS, $G_\infty=\liminf_{\bar\gamma\to\infty}(R_s^{(2)}(\vect{p}^\star)-R_s^{(2)}(\vect{p}_\mathrm{c}))$, satisfies
\begin{equation}
G_\infty\geq-\log_2(1-\rho_\mathrm{c}^2)-\log_2(1+\rho_{0,\mathrm{c}}).
\label{eq:gain}
\end{equation}
\end{proposition}
\begin{IEEEproof}
At high SNR, $R_s^{(2)}(\vect{p})=\log_2\bar\gamma+\log_2\big(\beta_1M(1-\rho^2)/\pi_1\big)+o(1)$ by~\eqref{eq:two}, since the offset term is constant. The double-null placement is feasible and gives $1-\rho^2=1$ and $\pi_1=1/(\beta_0NM)$, so $R_s^{(2)}(\vect{p}^\star)\geq R_s^{(2)}(\vect{p}_\mathrm{dn})$. For the compact array, $\pi_1\geq1/\lambda_1=1/(\beta_0NM(1+\rho_{0,\mathrm{c}}))$ by~\eqref{eq:pik} and~\eqref{eq:lam}. Taking the difference gives~\eqref{eq:gain}.
\end{IEEEproof}
In the setting of Section~\ref{sec:sim}, $\rho_\mathrm{c}=0.79$ and $\rho_{0,\mathrm{c}}=0.07$, so movement gains at least $1.3$\,bit/s/Hz. The bound grows without limit as $\rho_\mathrm{c}\to1$, but so does $D_\mathrm{dn}\propto1/|\dsin|$; the next result describes this regime.

\begin{proposition}[Small separation]
\label{prop:small}
Under LoS, for every $\vect{p}\in\mathcal{X}$,
\begin{equation}
1-\rho^2=\kappa^2\dsin^2\Var(\vect{p})+O\big((\kappa\dsin D)^4\big).
\label{eq:small}
\end{equation}
For even $M$, the maximum of $\Var(\vect{p})$ over $\mathcal{X}$ is attained by two compact clusters of $M/2$ elements at the two ends of the track and equals
\begin{equation}
\Var_{\max}=\frac{\big(D-(M/2-1)d_0\big)^2}{4}+\frac{d_0^2(M^2/4-1)}{12},
\label{eq:varmax}
\end{equation}
whereas the compact array has $\Var(\vect{p}_\mathrm{c})=d_0^2(M^2-1)/12$. Hence, as $\kappa\dsin D\to0$ and for equal costs $\pi_1$ at both placements, the gain of movement in the data SNR of~\eqref{eq:two} approaches the factor $\Var_{\max}/\Var(\vect{p}_\mathrm{c})$, which scales as $3D^2/(M^2d_0^2)$ for $D\gg Md_0$.
\end{proposition}
\begin{IEEEproof}
See Appendix~\ref{app:small}.
\end{IEEEproof}
Thus, for a nearby Eve, the best placement is a two-cluster interferometer with a long baseline; for $M=16$ and $D=16\lambda$, $\Var_{\max}/\Var(\vect{p}_\mathrm{c})=7.6$ ($2.9$\,bit/s/Hz) when $\kappa|\dsin|D\ll1$, i.e., $\Delta\theta\ll0.7^\circ$.

\section{Robust Design under Target-Angle Uncertainty}
\label{sec:robust}
This section relaxes the assumption of a perfectly known target direction, which in ISAC is obtained by sensing and is therefore imperfect. We quantify how the leakage of the data field grows with the angle error, and show that the same position spread that benefits secrecy sharpens the null and increases this sensitivity; we then propose a derivative-null design that suppresses the leakage to fourth order in the error and show that the sensing stream itself masks the residual leakage.
\subsection{Uncertainty Model and Robust Problem}
In ISAC, the target channel is learned from its echoes, so the BS knows the target direction only approximately, and robust secure-ISAC designs therefore protect an angular sector rather than a single direction~\cite{Ren2023robust,Hua2024secure,ZhaoH2026,Zhuang2026}. Here we characterize analytically how the sensitivity of the null depends on the placement. We model the target as a point scatterer with an LoS RIS--target channel $\vect{g}_\mathrm{E}(\theta)=\sqrt{\beta_1}\,\vect{a}(\theta)$, $\vect{a}(\theta)=\vect{b}(\vect{p};\sin\theta)$, where $\theta$ lies in the uncertainty sector $\Theta_\delta=[\hat\theta-\delta,\hat\theta+\delta]$ around the estimate $\hat\theta$. Bob's channel is assumed known, e.g., through uplink pilots. The worst-case counterpart of (P1) is
\begin{align}
(\mathrm{P2}):\ \max_{\vect{w}_1,\vect{w}_2,\Th,\vect{p}}\ \ & \min_{\theta\in\Theta_\delta}\Big[\log_2(1+\gamma_\mathrm{B})-\log_2(1+\gamma_\mathrm{E}(\theta))\Big]^+\nonumber\\
\mathrm{s.t.}\ \ & \|\vect{h}_\mathrm{E}(\theta)\|^2P_T(\theta)\geq\Gamma h_\mathrm{ref}\ \ \forall\theta\in\Theta_\delta,\nonumber\\
& \|\vect{w}_1\|^2+\|\vect{w}_2\|^2\leq P,\ \ \Th\in\mathcal{T},\ \ \vect{p}\in\mathcal{X}.
\label{eq:P2}
\end{align}
The echo SNR must now be met at every angle of the sector, and the secrecy rate is evaluated at the worst angle.

\subsection{Sensitivity of the Nominal Null}
The derivatives of the steering vector are $\dot{\vect{a}}(\theta)=j\kappa\cos\theta\,\mathrm{diag}(\vect{p})\vect{a}(\theta)$ and $\ddot{\vect{a}}(\theta)=-\mathrm{diag}\big(\kappa^2\cos^2\theta\,\vect{p}^2+j\kappa\sin\theta\,\vect{p}\big)\vect{a}(\theta)$, where $\vect{p}^2$ is taken entrywise. Let $\zeta(\theta)=|\vect{a}(\theta)^H\vect{d}_1|^2$ be the leakage per unit data energy and per unit $\beta_1$.
\begin{proposition}[Leakage under angle errors]
\label{prop:leak}
Let $\theta=\hat\theta+\nu$. (a) For the nominal ZF direction $\vect{d}_1\propto\vect{P}_{\vect{a}(\hat\theta)}^\perp\vect{g}_\mathrm{B}$,
\begin{equation}
\zeta(\theta)=\nu^2|\dot{\vect{a}}^H\vect{d}_1|^2+O(\nu^3)\leq \nu^2\kappa^2\cos^2\hat\theta\,M\Var(\vect{p})+O(\nu^3).
\label{eq:leaknom}
\end{equation}
(b) If $\vect{g}_\mathrm{B}\notin\mathrm{span}(\vect{Z})$, for the derivative-null direction
\begin{equation}
\vect{d}_1^\mathrm{rob}\propto\vect{P}_{\vect{Z}}^\perp\vect{g}_\mathrm{B},\qquad \vect{Z}=[\vect{a}(\hat\theta),\dot{\vect{a}}(\hat\theta)],
\label{eq:drob}
\end{equation}
$\zeta(\theta)=\tfrac{\nu^4}{4}|\ddot{\vect{a}}^H\vect{d}_1^\mathrm{rob}|^2+O(\nu^5)$. The price is a smaller data gain, $|\vect{g}_\mathrm{B}^H\vect{d}_1^\mathrm{rob}|^2=\|\vect{g}_\mathrm{B}\|^2(1-\rho^2-\dot\rho^2)$, where $\dot\rho=|\vect{g}_\mathrm{B}^H\dot{\vect{a}}_\perp|/(\|\vect{g}_\mathrm{B}\|\|\dot{\vect{a}}_\perp\|)$ and $\dot{\vect{a}}_\perp=\vect{P}_{\vect{a}(\hat\theta)}^\perp\dot{\vect{a}}(\hat\theta)$.
\end{proposition}
\begin{IEEEproof}
See Appendix~\ref{app:leak}.
\end{IEEEproof}
Proposition~\ref{prop:leak} reveals a tension specific to movable surfaces: the position variance that decorrelates Bob from Eve (Proposition~\ref{prop:small}) also sharpens the null at Eve and makes it more sensitive to angle errors, so a movable surface needs a robust null more than a compact one. The derivative-null direction removes the first-order sensitivity at the cost of one additional spatial degree of freedom, measured by $\dot\rho$, which the placement can again reduce.

\subsection{The Sensing Stream as Artificial Noise}
\begin{proposition}[Masking]
\label{prop:mask}
Consider any design with $\vect{g}_\mathrm{B}^H\vect{v}_2=0$ in which the sensing stream alone meets the echo constraint, i.e., $\|\vect{h}_\mathrm{E}(\theta)\|^2|\vect{g}_\mathrm{E}(\theta)^H\vect{v}_2|^2\geq\Gamma h_\mathrm{ref}$ for all $\theta\in\Theta_\delta$. Then, under LoS with resolvable equal-gain paths, for all $\theta\in\Theta_\delta$,
\begin{equation}
\gamma_\mathrm{E}(\theta)\leq\frac{\beta_1\mathcal{E}_1\zeta(\theta)}{\sigma^2+\Gamma\beta_0NM/\lambda_1}\leq\frac{\beta_1\mathcal{E}_1\zeta(\theta)}{\sigma^2+\Gamma/2},
\label{eq:mask}
\end{equation}
with $\sigma^2+\Gamma$ in the denominator for an isotropic incident subspace.
\end{proposition}
\begin{IEEEproof}
The numerator of $\gamma_\mathrm{E}(\theta)$ in~\eqref{eq:sinr} is $\beta_1\mathcal{E}_1\zeta(\theta)$, and the denominator is $\sigma^2+|\vect{g}_\mathrm{E}(\theta)^H\vect{v}_2|^2\geq\sigma^2+\Gamma h_\mathrm{ref}/\|\vect{h}_\mathrm{E}(\theta)\|^2$, where $\|\vect{h}_\mathrm{E}(\theta)\|^2\leq\lambda_1\beta_1M$ by Lemma~\ref{lem:ret}, $\lambda_1\leq2\beta_0NM$, and $\|\vect{h}_\mathrm{E}\|^2\leq\beta_0\beta_1NM^2$ in the isotropic case. Bob's SINR is unaffected because $\vect{g}_\mathrm{B}^H\vect{v}_2=0$.
\end{IEEEproof}
Proposition~\ref{prop:mask} shows that, in secure ISAC, the sensing requirement is also a secrecy guarantee: every unit of illumination that the echo requires is Gaussian noise for Eve, and it divides the residual leakage by at least $1+\bar\Gamma/2$, and by $1+\bar\Gamma$ for an isotropic incident subspace.  

\subsection{Robust Design}
Combining Propositions~\ref{prop:leak} and~\ref{prop:mask}, we use the data direction~\eqref{eq:drob} and the nominal sensing direction $\vect{d}_2\propto\vect{P}^\perp_{\vect{g}_\mathrm{B}}\vect{a}(\hat\theta)$, The fields span $\Sp_\mathrm{rob}=\mathrm{span}\{\vect{d}_1^\mathrm{rob},\vect{d}_2\}\neq\Sp$, but Corollary~\ref{cor:sub} applies with $\vect{Q}_0$ a basis of $\Sp_\mathrm{rob}$, the costs follow from~\eqref{eq:pik}, and the sensing energy is chosen for the worst angle of the sector,
\begin{equation}
\mathcal{E}_2(\phi)=\frac{\Gamma h_\mathrm{ref}}{\min_{\theta\in\Theta_\delta}\|\vect{h}_\mathrm{E}(\theta,\phi)\|^2\beta_1|\vect{a}(\theta)^H\vect{d}_2|^2},
\label{eq:E2rob}
\end{equation}
with the return gain $\|\vect{h}_\mathrm{E}(\theta,\phi)\|^2=\beta_1\sum_i\lambda_i|\vect{a}(\theta)^H\vect{f}_i(\phi)|^2$ of Lemma~\ref{lem:ret}. With $\mathcal{E}_1(\phi)=(P-\pi_2(\phi)\mathcal{E}_2(\phi))/\pi_1(\phi)$, the worst-case secrecy rate is
\begin{align}
R_s^\mathrm{rob}=\max_{\phi}\Big[&\log_2\big(1+|\vect{g}_\mathrm{B}^H\vect{d}_1^\mathrm{rob}|^2\mathcal{E}_1/\sigma^2\big)\nonumber\\
&-\max_{\theta\in\Theta_\delta}\log_2\Big(1+\frac{\beta_1\mathcal{E}_1\zeta(\theta)}{\sigma^2+\beta_1\mathcal{E}_2|\vect{a}(\theta)^H\vect{d}_2|^2}\Big)\Big],
\label{eq:Rrob}
\end{align}
evaluated on a grid of $N_\theta$ angles; the data stream's own illumination adds margin to the sensing constraint. The nominal design uses $\vect{d}_1$ of~\eqref{eq:zf} in~\eqref{eq:Rrob} and is preferable for small $\delta$; the BS can pick the better of the two. The position search of Section~\ref{sec:alg} applies unchanged with~\eqref{eq:Rrob} as the objective.

\section{Joint Placement and Scattering Design}
\label{sec:alg}
This section turns the analysis into an algorithm. For given element positions, every remaining variable follows in closed form from Sections~\ref{sec:real}--\ref{sec:robust}, so the design reduces to a search over the positions, for which we propose a monotone coordinate search, establish its convergence and complexity, and describe the benchmark schemes used in Section~\ref{sec:sim}.
\subsection{Closed-Form Design for Given Positions}
For given positions, the design is closed-form: 1) the rank-two SVD of $\vect{H}$, via the $2\times2$ matrix $\vect{D}_0^{1/2}\vect{B}^H\vect{B}\vect{D}_0^{1/2}$; 2) the directions~\eqref{eq:zf} (or~\eqref{eq:drob}) and a basis $\vect{Q}_0$ of their span; 3) the SVD of $\vect{E}^T\vect{Q}_0$, the costs $\pi_k(\phi)$, and the return gain $\|\vect{h}_\mathrm{E}(\phi)\|^2$ (Lemmas~\ref{lem:ret},~\ref{lem:phase}); 4) $\phi$ and the energies by Theorem~\ref{thm:two} (or~\eqref{eq:Rrob}); 5) $\vect{t}_k=\vect{F}(\phi)^H\vect{v}_k$ and $\vect{w}_k=\vect{W}\bm{\Lambda}^{-1/2}\vect{t}_k$ (Lemma~\ref{lem:inc}); 6) $\Th$ by Theorem~\ref{thm:real}. Steps 1)--4) evaluate the rate during the position search; steps 5)--6) are executed once.

\subsection{Position Optimization}
The rate is highly oscillatory in each position, so gradient methods stall at local maxima. We therefore use a coordinate search: element $m$ moves to the best point of a grid of step $\delta_x$ over the interval allowed by its neighbors,
\begin{equation}
\mathcal{I}_m=[p_{m-1}+d_0,\ p_{m+1}-d_0],
\label{eq:Im}
\end{equation}
with $p_0=-d_0$, $p_{M+1}=D+d_0$, and the current position kept as a candidate. The search is started from the compact array, the double-null placement (if $D\geq D_\mathrm{dn}$), and a random placement, and the best result is kept (Algorithm~\ref{alg:main}).

\begin{algorithm}[t]
\caption{Movable BD-RIS design for secure ISAC}
\label{alg:main}
\begin{algorithmic}[1]
\Require channels, $P$, $\Gamma$, initial placements $\{\vect{p}^{(0)}\}\subset\mathcal{X}$, grid step $\delta_x$, tolerance $\epsilon$
\For{each initial placement $\vect{p}^{(0)}$; $t\leftarrow0$, $\vect{p}\leftarrow\vect{p}^{(0)}$}
\Repeat
\For{$m=1,\dots,M$}
\State $p_m\leftarrow\arg\max_{x\in\mathrm{grid}(\mathcal{I}_m)\cup\{p_m\}}R_s^{(2)}(\vect{p})$
\EndFor
\State $t\leftarrow t+1$, $R^{(t)}\leftarrow R_s^{(2)}(\vect{p})$
\Until{$R^{(t)}-R^{(t-1)}\leq\epsilon$}
\EndFor
\State keep the best placement $\vect{p}$
\State ZF fields $\vect{v}_1,\vect{v}_2$ (Theorem~\ref{thm:two}); $\vect{t}_k=\vect{F}^H\vect{v}_k$; $\vect{w}_k=\vect{W}\bm{\Lambda}^{-1/2}\vect{t}_k$
\State $\Th$ from Theorem~\ref{thm:real} with $\vect{U}=[\vect{H}\vect{w}_1,\vect{H}\vect{w}_2]$, $\vect{V}=[\vect{v}_1,\vect{v}_2]$
\Ensure $\vect{p}$, $\Th$, $\vect{w}_1$, $\vect{w}_2$
\end{algorithmic}
\end{algorithm}

The merit function $\tilde R(\vect{p})$ used in the search equals $R_s^{(2)}$ when the placement is feasible and the negative power deficit $\max_\phi(P-\pi_2(\phi)\mathcal{E}_2(\phi))/P$ otherwise, which guides the search toward feasibility.

\emph{Remark:} Steps 1)--4) evaluate the rate during the position search; steps 5)--6) are executed once. No alternation between the positions and the remaining variables is needed: for every candidate placement, steps 1)--4) return the exact optimum of the ZF-constrained inner problem, so the position search operates directly on the best-response rate $R_s^{(2)}(\vect{p})$, unlike the movable D-RIS benchmark.

\subsection{Convergence and Complexity}
\begin{proposition}[Convergence]
\label{prop:conv}
For each initial placement, the merit sequence $\{\tilde R^{(t)}\}$ generated by Algorithm~\ref{alg:main} is nondecreasing and bounded above by $R_{\max}=\log_2\big(1+\lambda_1^{\max}\beta_1M(\sum_{l}|\alpha_{\mathrm{B},l}|)^2\bar\gamma\big)$. Hence, for $\epsilon>0$, the algorithm terminates after at most $\lceil(R_{\max}-\tilde R^{(0)})/\epsilon\rceil+1$ sweeps; for $\epsilon=0$ and $d_0/\delta_x\in\mathbb{Q}$, it terminates at a placement that is coordinatewise optimal on the grid, provided the current position is kept whenever it ties.
\end{proposition}
\begin{IEEEproof}
Each coordinate update keeps the current position as a candidate, so $\tilde R$ does not decrease. By~\eqref{eq:two} and Lemma~\ref{lem:inc}, $R_s^{(2)}\leq\log_2(1+\lambda_1\|\vect{g}_\mathrm{B}\|^2P/\sigma^2)$, with $\lambda_1\leq\lambda_1^{\max}$ and $\|\vect{g}_\mathrm{B}\|^2\leq\beta_1M(\sum_l|\alpha_{\mathrm{B},l}|)^2$ for every placement, which gives $R_{\max}$ (the deficit is negative). Every sweep that does not stop the algorithm increases $\tilde R$ by more than $\epsilon$, which gives the bound on the number of sweeps. For $\epsilon=0$ and a rational ratio $d_0/\delta_x$, all positions visited after the first sweep lie on a common lattice, so the reachable placements form a finite set, and a nondecreasing sequence on a finite set becomes constant at a placement that no single grid move improves.
\end{IEEEproof}
One candidate position costs $\mathcal{O}(MN+N_\phi)$ (channel update, rank-two SVD, projections, and phase grid), so a sweep over the $M$ elements with at most $D/\delta_x$ candidates each costs $\mathcal{O}\big(M(D/\delta_x)(MN+N_\phi)\big)$. With $T$ sweeps and $N_\mathrm{s}$ initial placements, followed by the one-time synthesis of $\Th$ in $\mathcal{O}(M^3)$, the total complexity of Algorithm~\ref{alg:main} is
\begin{equation}
\mathcal{O}\big(N_\mathrm{s}TM(D/\delta_x)(MN+N_\phi)+M^3\big),
\label{eq:cx}
\end{equation}
which is polynomial in all parameters and, for a fixed track, quadratic in $M$; the robust objective adds $\mathcal{O}(MN_\theta)$ per candidate. In the simulations, $T\leq8$ and $N_\mathrm{s}=3$, and one sweep takes about $0.1$\,s for $M=16$. By contrast, an alternating optimization over $\mathcal{T}$ would require an $\mathcal{O}(M^3)$ projection onto the symmetric unitary manifold at every inner iteration and every candidate.

\begin{table}[t]
\centering
\caption{Simulation Parameters}
\label{tab:par}
\renewcommand{\arraystretch}{1.12}
\begin{tabular}{l l}
\hline
Parameter & Value\\
\hline
BS antennas $N$, RIS elements $M$ & $8$, $16$\\
Track length $D$, minimum spacing $d_0$ & $16\lambda$, $\lambda/2$\\
BS--RIS paths (at the RIS) & $-40^\circ$, $-10^\circ$, equal gains\\
Bob, Eve directions & $30^\circ$, $30^\circ-\Delta\theta$, $\Delta\theta=3^\circ$\\
Rician factor $K_R$, scattered paths $L$ & $10$\,dB, $4$\\
Reference SNR $\bar\gamma$, sensing requirement $\bar\Gamma$ & $0$\,dB, $10$\,dB\\
Position grid $\delta_x$, tolerance $\epsilon$ & $\lambda/20$, $10^{-4}$\\
Phase grids $N_\phi$, $N_\vartheta$, leakage grid $N_\ell$ & $64$, $32$, $2001$\\
Sector grid $N_\theta$ (Section~\ref{sec:robust}) & $21$\\
\hline
\end{tabular}
\end{table}
\subsection{Benchmarks}
We define the following benchmarks: \emph{1) Fixed BD-RIS:} Theorem~\ref{thm:two} at the compact array. \emph{2) Single-stream BD-RIS (fixed/movable):} Theorem~\ref{thm:one} with a search over $N_\ell$ leakage values, at the compact array or with optimized positions. \emph{3) D-RIS (fixed/movable):} since a D-RIS cannot realize the ZF fields (Proposition~\ref{prop:dris}), ZF is applied to the effective channels $\vect{h}_X^H=\vect{g}_X^H\Th\vect{H}$. The precoders $\vect{w}_1\propto\vect{P}^\perp_{\vect{h}_\mathrm{E}}\vect{h}_\mathrm{B}$ and $\vect{w}_2\propto\vect{P}^\perp_{\vect{h}_\mathrm{B}}\vect{h}_\mathrm{E}$ give $\gamma_\mathrm{E}=0$ and the rate $\log_2\big(1+\|\vect{h}_\mathrm{B}\|^2(1-\rho_h^2)(P-P_2)/\sigma^2\big)$, with $P_2=\Gamma h_\mathrm{ref}/(\|\vect{h}_\mathrm{E}\|^4(1-\rho_h^2))$ from~\eqref{eq:sens}, since the echo returns over the same effective channel, and $\rho_h$ the correlation of the effective channels. The phases are optimized by coordinate ascent over a grid of $N_\vartheta$ points, and for the movable D-RIS the positions and the phases are optimized alternately.

\section{Numerical Results}
\label{sec:sim}
This section validates the analysis and quantifies the gains of moving, interconnecting, and robustifying the surface under Rician fading, against fixed, diagonal, and single-stream benchmarks.
\subsection{Simulation Setup}
\begin{figure}[t]
\centering
\includegraphics[width=\columnwidth]{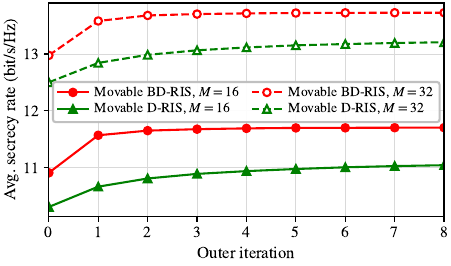}
\caption{Convergence of the position search from a random placement ($\bar\Gamma=10$\,dB; solid: $M=16$, $D=16\lambda$; dashed: $M=32$, $D=32\lambda$).}
\label{fig:conv}
\end{figure}

The parameters are listed in Table~\ref{tab:par}. The BS-side directions of the two BS--RIS paths make them resolvable, $\vect{A}^H\vect{A}=N\vect{I}_2$, and $\Delta\theta=3^\circ$ gives $D_\mathrm{dn}=15.6\lambda$. Scattered paths have directions uniform in $[-90^\circ,90^\circ]$, and $\beta_0=\beta_1=1$, so $\bar\gamma$ is the SNR before the cascaded array gain $NM^2$. Results are averaged over $100$ realizations ($60$ in Fig.~\ref{fig:M}, $50$ in Fig.~\ref{fig:conv}), with standard errors below $0.15$\,bit/s/Hz except near the feasibility limit of the fixed designs (at most $0.34$\,bit/s/Hz). The six schemes of Section~\ref{sec:alg} are compared.

\begin{figure*}[t]
\centering
\includegraphics[width=\textwidth]{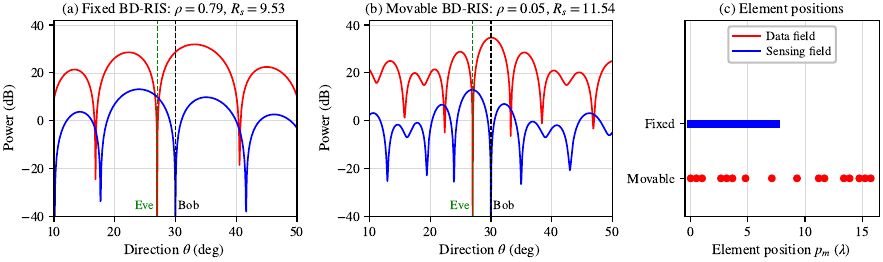}
\caption{Radiation patterns of the data field $|\vect{a}(\theta)^H\vect{v}_1|^2$ and the sensing field $|\vect{a}(\theta)^H\vect{v}_2|^2$ under LoS ($\Delta\theta=3^\circ$, $\bar\Gamma=10$\,dB, $R_s$ in bit/s/Hz). (a) Fixed BD-RIS. (b) Movable BD-RIS (Algorithm~\ref{alg:main}). (c) Element positions.}
\label{fig:beam}
\end{figure*}

\subsection{Field Patterns}
Fig.~\ref{fig:beam} illustrates Theorem~\ref{thm:two} under LoS. Both surfaces null the data at Eve and the sensing at Bob; the synthesized $\Th$ is symmetric, unitary, satisfies $\Th\vect{U}=\vect{V}$ to within $3\times10^{-14}$, and meets~\eqref{eq:sens} with equality. With the compact array, Bob and Eve share the main lobe ($\rho=0.79$), so Bob sits on the flank of the data null. The movable surface spreads its elements, as Proposition~\ref{prop:small} predicts, so that the null and the peak of each field are $3^\circ$ apart ($\rho=0.05$), and $R_s$ increases from $9.53$ to $11.54$\,bit/s/Hz.

\subsection{Convergence}
In Fig.~\ref{fig:conv}, the rate increases monotonically, as Proposition~\ref{prop:conv} guarantees, and reaches $99.8\%$ of its final value within three outer iterations. The movable D-RIS converges more slowly and settles $0.7$\,bit/s/Hz lower. Doubling $M$ and $D$ adds $2.0$\,bit/s/Hz, consistent with the $M^2$ scaling of~\eqref{eq:dn}.

\begin{figure*}[t]
\centering
\includegraphics[width=0.97\textwidth]{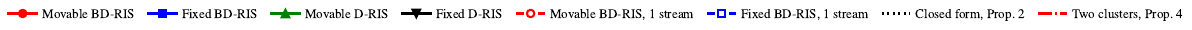}\\[1pt]
\begin{minipage}[t]{0.32\textwidth}
\centering\includegraphics[width=\textwidth]{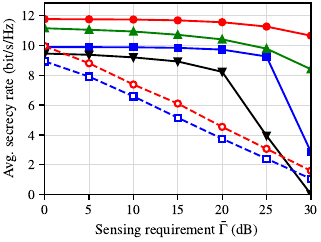}
\caption{Secrecy rate vs.\ sensing requirement $\bar\Gamma$.}
\label{fig:gamma}
\end{minipage}\hfill
\begin{minipage}[t]{0.32\textwidth}
\centering\includegraphics[width=\textwidth]{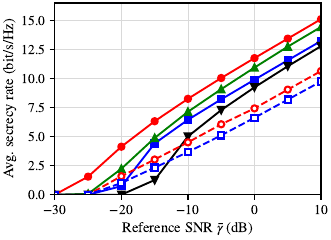}
\caption{Secrecy rate vs.\ reference SNR $\bar\gamma$ ($\bar\Gamma=10$\,dB).}
\label{fig:snr}
\end{minipage}\hfill
\begin{minipage}[t]{0.32\textwidth}
\centering\includegraphics[width=\textwidth]{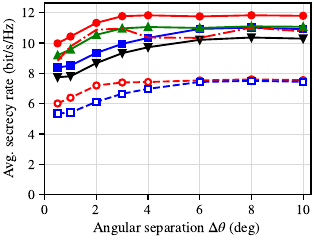}
\caption{Secrecy rate vs.\ Bob--Eve separation $\Delta\theta$.}
\label{fig:angle}
\end{minipage}
\end{figure*}

\subsection{Impact of the Sensing Requirement}
Fig.~\ref{fig:gamma} shows the central trade-off. To quantify the ZF restriction of Theorem~\ref{thm:two}, we also solved the unrestricted two-stream problem over all field pairs in $\Sp$ at the optimized positions (sequential quadratic programming over the four complex coefficients for each of $16$ frame phases, from the ZF design and six random starts). It exceeds~\eqref{eq:two} by at most $0.04$\,bit/s/Hz on average and $0.06$ in the worst realization over Figs.~\ref{fig:gamma} and~\ref{fig:snr}, with the largest gaps at $\bar\Gamma=30$\,dB and $\bar\gamma=-20$\,dB, where leakage is cheapest; the curves coincide within the line width, so only the ZF design is shown. The dual-stream movable BD-RIS loses only $1.1$\,bit/s/Hz when $\bar\Gamma$ grows from $0$ to $30$\,dB, in agreement with the SNR offset in~\eqref{eq:twoiso}, whereas the single-stream design drops by $8.3$\,bit/s/Hz, as Theorem~\ref{thm:one} predicts. Two streams outperform one for every $\bar\Gamma\geq0$\,dB, a wider range than the sufficient condition of Corollary~\ref{cor:win}. At $\bar\Gamma=10$\,dB, movement adds $1.9$\,bit/s/Hz to the BD-RIS and $1.7$\,bit/s/Hz to the D-RIS, and interconnection adds $0.8$\,bit/s/Hz at optimized positions; the gap to the movable D-RIS widens to $2.3$\,bit/s/Hz at $\bar\Gamma=30$\,dB, because the D-RIS can neither steer the sensing field independently (Proposition~\ref{prop:dris}) nor choose its return gain. The fixed surfaces collapse at $\bar\Gamma=30$\,dB, because $\rho\approx0.8$ makes the sensing stream $1/(1-\rho^2)\approx2.7$ times costlier (Theorem~\ref{thm:two}).

\subsection{Impact of the Reference SNR}
Fig.~\ref{fig:snr} shows the secrecy rate versus the reference SNR. At high SNR, the gains become constant offsets: at $\bar\gamma=10$\,dB, movement adds $1.9$\,bit/s/Hz to the BD-RIS, above the LoS lower bound of $1.3$\,bit/s/Hz in Proposition~\ref{prop:gain}, since the position search also exploits the fading, and the dual-stream design exceeds the single-stream design by $4.5$\,bit/s/Hz, well above the guaranteed $\log_2((1+\bar\Gamma/2)/2)=1.6$\,bit/s/Hz of Corollary~\ref{cor:win}. At low SNR, feasibility dominates: at $\bar\gamma=-20$\,dB, the fixed single-stream design ($1.0$\,bit/s/Hz) outperforms the fixed dual-stream design ($0.7$\,bit/s/Hz), as anticipated after Corollary~\ref{cor:win}. The movable BD-RIS, however, remains feasible down to $\bar\gamma=-25$\,dB and attains $4.1$\,bit/s/Hz at $-20$\,dB, $1.9$\,bit/s/Hz more than the best benchmark, because decorrelation reduces the sensing power.

\subsection{Impact of the Angular Separation}
In Fig.~\ref{fig:angle}, the benefit of movement concentrates on nearby eavesdroppers: at $\Delta\theta=1^\circ$, movement adds $1.9$\,bit/s/Hz to the BD-RIS ($10.4$ vs.\ $8.5$\,bit/s/Hz), and beyond $\Delta\theta\approx6^\circ$, where the compact array already decorrelates Bob from Eve, the gain drops to $0.8$\,bit/s/Hz, which stems from $\rho_0$ and the fading. The two-cluster placement of Proposition~\ref{prop:small} (dash-dotted) captures $60$--$80\%$ of the gain for $\Delta\theta\leq3^\circ$ without any search, but falls back to the fixed-array level for $\Delta\theta\geq4^\circ$, where $\kappa|\dsin|D>4$ and the small-separation regime ends; the coordinate search is needed there. The single-stream curves remain $4.0$--$4.4$\,bit/s/Hz below their dual-stream counterparts, more than $\log_2(1+\bar\Gamma)$: the single beam is aimed at Bob, so the frame returns little of the echo, and the illumination, which is information-bearing, must grow accordingly. With two streams, the sensing field is aimed at the target, and its echo returns through a frame that the design can choose.

\begin{figure*}[t]
\centering
\includegraphics[width=0.97\textwidth]{fig_legend.pdf}\\[1pt]
\begin{minipage}[t]{0.32\textwidth}
\centering\includegraphics[width=\textwidth]{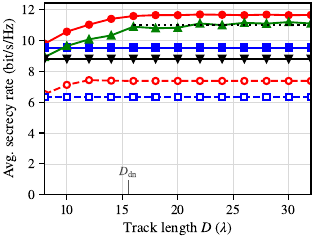}
\caption{Secrecy rate vs.\ track length $D$ under LoS; the tick marks $D_\mathrm{dn}$.}
\label{fig:track}
\end{minipage}\hfill
\begin{minipage}[t]{0.32\textwidth}
\centering\includegraphics[width=\textwidth]{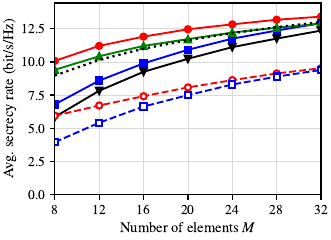}
\caption{Secrecy rate vs.\ number of elements $M$ ($D=20\lambda$).}
\label{fig:M}
\end{minipage}\hfill
\begin{minipage}[t]{0.32\textwidth}
\centering\includegraphics[width=\textwidth]{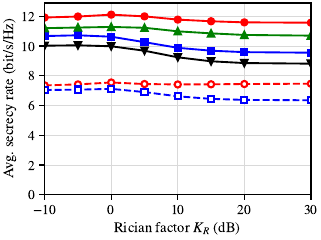}
\caption{Secrecy rate vs.\ Rician factor $K_R$ of the RIS--user links.}
\label{fig:K}
\end{minipage}
\end{figure*}

\subsection{Impact of the Track Length}
Fig.~\ref{fig:track} validates Proposition~\ref{prop:dn} under LoS. The double-null placement attains the closed form~\eqref{eq:dn}, namely $11.0$\,bit/s/Hz, exactly for $D\geq D_\mathrm{dn}$: its incident subspace is isotropic, so the return gain equals $h_\mathrm{ref}$ (Lemma~\ref{lem:ret}). Algorithm~\ref{alg:main} exceeds it by up to $0.7$\,bit/s/Hz by sending the data stream through a correlated incident pair ($\rho_0>0$). For $D<D_\mathrm{dn}$, the residual correlation bounded by Proposition~\ref{prop:dn} costs $1.2$\,bit/s/Hz at $D=8\lambda$, where the converse gives $\rho\geq\cos(\pi|\dsin|D/\lambda)=0.40$. The rate saturates beyond $D_\mathrm{dn}$, so a track of about $D_\mathrm{dn}$ suffices in practice, and the single-stream design saturates $4.3$\,bit/s/Hz lower.

\subsection{Impact of the Number of Elements}
Fig.~\ref{fig:M} varies $M$ on a track of fixed length $D=20\lambda$, which satisfies $D\geq D_\mathrm{dn}$ for every $M$ shown. The LoS closed form~\eqref{eq:dn} grows by $2.0$\,bit/s/Hz per doubling of $M$, reflecting the $M^2$ array gain, and the movable BD-RIS exceeds it at every $M$ by exploiting $\rho_0>0$ and the fading. The fixed BD-RIS grows faster, by $3.0$\,bit/s/Hz per doubling, because its aperture, and hence its decorrelation, grows with $M$; the gain of movement therefore shrinks from $3.3$\,bit/s/Hz at $M=8$ to $0.5$\,bit/s/Hz at $M=32$, where the compact array already occupies $15.5\lambda$ of the track. Movement thus substitutes for elements: the movable BD-RIS with $M=8$ ($10.1$\,bit/s/Hz) outperforms the fixed BD-RIS with $M=16$ ($9.9$\,bit/s/Hz), and with $M=16$ it outperforms the fixed BD-RIS with $M=24$. Interconnection adds $0.5$--$0.7$\,bit/s/Hz at every $M$, and the single-stream designs remain at least $3.9$\,bit/s/Hz below.

\subsection{Impact of the Propagation Environment}
In Fig.~\ref{fig:K}, scattering decorrelates Bob from Eve even at the compact array, so the gain of movement shrinks from $2.0$\,bit/s/Hz at $K_R=30$\,dB to $1.3$\,bit/s/Hz at $K_R=-10$\,dB: movement matters most in LoS-dominated links. Interconnection adds $0.7$--$0.9$\,bit/s/Hz for every $K_R$.

\begin{figure*}[t]
\centering
\begin{minipage}[t]{0.32\textwidth}
\centering\includegraphics[width=\textwidth]{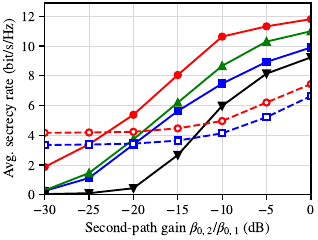}
\caption{Secrecy rate vs.\ gain of the second BS--RIS path, i.e., the strength of the second incident mode (legend as in Fig.~\ref{fig:gamma}).}
\label{fig:b2}
\end{minipage}\hfill
\begin{minipage}[t]{0.66\textwidth}
\centering\includegraphics[width=\textwidth]{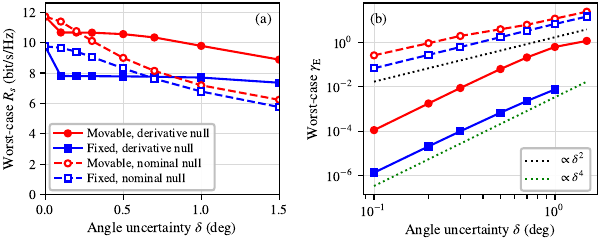}
\caption{Robustness to target-angle uncertainty: (a) worst-case secrecy rate over $\Theta_\delta$; (b) worst-case Eve SINR (geometric mean over realizations) with reference slopes.}
\label{fig:robust}
\end{minipage}
\end{figure*}

\subsection{Impact of the Rank of the BS--RIS Link}
Fig.~\ref{fig:b2} confirms the role of the rank of the BS--RIS link (Lemmas~\ref{lem:inc},~\ref{lem:ret}): as the second path weakens, $\lambda_2\approx\beta_{0,2}NM(1-\rho_0^2)$ vanishes and the dual-stream designs, which need both incident modes, collapse, whereas the single-stream designs lose only the second mode's share of the return gain. The movable BD-RIS keeps $10.6$\,bit/s/Hz down to $-10$\,dB and stays above its single-stream counterpart down to $-20$\,dB, because it places the elements to drive $\rho_0$ toward zero and thus maximize $\lambda_2$; the fixed BD-RIS crosses its single-stream counterpart near $-20$\,dB as well, but at a much lower rate. The BS should therefore switch to one stream when the second path is more than about $20$\,dB weaker than the first.

\subsection{Robustness to Target-Angle Uncertainty}
Fig.~\ref{fig:robust} evaluates Section~\ref{sec:robust} with an LoS target anywhere in $\Theta_\delta$, a Rician channel for Bob, and positions optimized for~\eqref{eq:Rrob}. With the nominal null, the movable BD-RIS loses $4.5$\,bit/s/Hz between $\delta=0$ and $\delta=1^\circ$, whereas the fixed BD-RIS loses $3.0$\,bit/s/Hz, so the advantage of movement shrinks to $0.4$\,bit/s/Hz. Fig.~\ref{fig:robust}(b) explains why: at $\delta=0.1^\circ$, the worst-case Eve SINR of the movable surface is $5.8$\,dB above that of the fixed surface, because its larger position variance sharpens the null, as~\eqref{eq:leaknom} predicts. The derivative-null design removes this fragility: its leakage grows by $37.5$\,dB per decade of $\delta$ for both surfaces up to $\delta=1^\circ$, i.e., as $\delta^4$, whereas the nominal null of the fixed surface grows by $19.9$\,dB per decade, i.e., as $\delta^2$, in agreement with Proposition~\ref{prop:leak}. The extra null costs $1.0$\,bit/s/Hz for the movable surface but $2.0$\,bit/s/Hz for the fixed one, whose compact array cannot separate Bob from the derivative direction ($\dot\rho$ is large). The movable derivative-null design keeps $10.6$\,bit/s/Hz up to $\delta=0.5^\circ$ and $9.8$\,bit/s/Hz at $\delta=1^\circ$, i.e., $2.6$\,bit/s/Hz more than the nominal null and $2.1$\,bit/s/Hz more than the best fixed design. With the masking of Proposition~\ref{prop:mask}, its worst-case Eve SINR stays below $0.7$\,dB throughout. Three checks complete the picture. First, re-evaluating the final designs on $2001$ angles of $\Theta_\delta$ instead of the $N_\theta=21$ design points leaves the worst-case rates unchanged and the echo constraint satisfied (margin $\geq0.1\%$), as the worst angle lies at the sector edge. Second, jointly optimizing $(\mathcal{E}_1,\mathcal{E}_2)$ per frame phase, counting the illumination of both streams and allowing surplus AN, barely changes the derivative-null design ($+0.4$\,bit/s/Hz at $\delta=1^\circ$) but lifts the nominal null to $11.4$, $10.6$, and $9.7$\,bit/s/Hz at $\delta=0.2^\circ$, $0.5^\circ$, and $1^\circ$: surplus AN substitutes for the extra null at small $\delta$, and the derivative null keeps its advantage ($10.1$ vs.\ $9.7$\,bit/s/Hz) only at $\delta=1^\circ$. Third, with $L=4$ unknown scattered paths at Eve ($K_R=10$\,dB), the fourth-order law no longer applies; both designs settle between $9.3$ and $10.7$\,bit/s/Hz, the derivative null being flatter ($9.8$\,bit/s/Hz for all $\delta$) and better for $\delta\geq0.5^\circ$.

\section{Conclusion}
\label{sec:concl}
In this paper, we studied secure ISAC with a movable BD-RIS when the sensed target is itself an eavesdropper and the echo returns to the BS through the surface. We proved that a lossless reciprocal BD-RIS realizes $K$ prescribed fields if and only if their Gram matrices coincide and their bilinear Gram matrix is symmetric. Through a rank-two link, the surface can therefore null the data at the target and the sensing signal at the user, so that sensing costs only an SNR offset instead of an unavoidable rate loss, and decorrelating the user from the target, which element movement provides, serves both tasks. The resulting closed-form ZF design lies within $0.06$\,bit/s/Hz of the unrestricted two-stream optimum in all tested settings, and, since all variables other than the positions follow in closed form, the joint design requires no alternating optimization. We further showed that the position variance that drives the gain of movement also sharpens the nulls, and that the derivative-null design, together with the masking provided by the sensing stream, protects the secrecy rate against target-angle errors. Finally, the dual-stream operation relies on a rank-two BS--RIS link and on an idealized lossless, fully connected network; with a single dominant path, a single stream remains the better choice.

\appendices
\section{Proof of Theorem~\ref{thm:real}}
\label{app:real}
\emph{Step 1 (Takagi factorization).} $\Th\in\mathcal{T}$ if and only if $\Th=\vect{Q}\vect{Q}^T$ with $\vect{Q}$ unitary: the product is symmetric and unitary, and conversely the Takagi factorization $\Th=\vect{Y}\bm{\Sigma}\vect{Y}^T$ of a complex symmetric matrix~\cite{HornJohnson2013} has $\bm{\Sigma}=\vect{I}_M$ when $\Th$ is unitary.

\emph{Step 2 (Reformulation).} With $\Th=\vect{Q}\vect{Q}^T$ and $\widetilde{\vect{U}}=\vect{Q}^T\vect{U}$, i.e., $\vect{Q}\widetilde{\vect{U}}^*=\vect{U}^*$, the equation $\Th\vect{U}=\vect{V}$ reads $\vect{Q}\widetilde{\vect{U}}=\vect{V}$. Hence it is solvable in $\mathcal{T}$ if and only if there exist $\widetilde{\vect{U}}\in\mathbb{C}^{M\times K}$ and a unitary $\vect{Q}$ with
\begin{equation}
\vect{Q}[\widetilde{\vect{U}},\widetilde{\vect{U}}^*]=[\vect{V},\vect{U}^*].
\label{eq:app1}
\end{equation}

\emph{Step 3 (Unitary extension).} For $\vect{Z},\vect{Y}\in\mathbb{C}^{M\times n}$, a unitary $\vect{Q}$ with $\vect{Q}\vect{Z}=\vect{Y}$ exists if and only if $\vect{Z}^H\vect{Z}=\vect{Y}^H\vect{Y}$. This is standard~\cite{HornJohnson2013}: with $\vect{Z}^H\vect{Z}=\vect{C}_+\bm{\Delta}_+\vect{C}_+^H$ (nonzero eigenvalues), $\vect{Z}\vect{C}_+\bm{\Delta}_+^{-1/2}$ and $\vect{Y}\vect{C}_+\bm{\Delta}_+^{-1/2}$ have orthonormal columns, and mapping their orthonormal completions onto each other gives $\vect{Q}$.

\emph{Step 4 (Gram matrices).} Let $\vect{G}=\vect{U}^H\vect{U}$ and $\vect{S}=\vect{U}^T\vect{V}$. Using (i) and (ii), $\vect{S}^T=\vect{S}$ and
\begin{equation}
[\vect{V},\vect{U}^*]^H[\vect{V},\vect{U}^*]=\begin{bmatrix}\vect{V}^H\vect{V}&\vect{V}^H\vect{U}^*\\ \vect{U}^T\vect{V}&\vect{U}^T\vect{U}^*\end{bmatrix}=\begin{bmatrix}\vect{G}&\vect{S}^*\\ \vect{S}&\vect{G}^*\end{bmatrix}\triangleq\vect{K},
\label{eq:app2}
\end{equation}
since $\vect{V}^H\vect{U}^*=\vect{S}^*$ and $\vect{U}^T\vect{U}^*=\vect{G}^*$. The Gram matrix of $[\widetilde{\vect{U}},\widetilde{\vect{U}}^*]$ has the same block form with $\widetilde{\vect{U}}^H\widetilde{\vect{U}}$ and the symmetric $\widetilde{\vect{U}}^T\widetilde{\vect{U}}$ in place of $\vect{G}$ and $\vect{S}$, so by Step 3, \eqref{eq:app1} is solvable if and only if there exists $\widetilde{\vect{U}}$ with
\begin{equation}
\widetilde{\vect{U}}^H\widetilde{\vect{U}}=\vect{G},\qquad \widetilde{\vect{U}}^T\widetilde{\vect{U}}=\vect{S}.
\label{eq:app3}
\end{equation}
\emph{Step 5 (Real embedding).} Write $\tilde{\vect{u}}_i=\bm{\xi}_i+j\bm{\eta}_i$ with $\bm{\xi}_i,\bm{\eta}_i\in\mathbb{R}^M$. Then
\begin{align}
\tilde{\vect{u}}_i^H\tilde{\vect{u}}_k&=\bm{\xi}_i^T\bm{\xi}_k+\bm{\eta}_i^T\bm{\eta}_k+j(\bm{\xi}_i^T\bm{\eta}_k-\bm{\eta}_i^T\bm{\xi}_k),\nonumber\\
\tilde{\vect{u}}_i^T\tilde{\vect{u}}_k&=\bm{\xi}_i^T\bm{\xi}_k-\bm{\eta}_i^T\bm{\eta}_k+j(\bm{\xi}_i^T\bm{\eta}_k+\bm{\eta}_i^T\bm{\xi}_k),
\end{align}
so~\eqref{eq:app3} prescribes all inner products of the $2K$ real vectors $(\bm{\xi}_1,\dots,\bm{\xi}_K,\bm{\eta}_1,\dots,\bm{\eta}_K)$:
\begin{align}
\bm{\xi}_i^T\bm{\xi}_k&=\tfrac12\Re\{G_{ik}+S_{ik}\},\quad \bm{\eta}_i^T\bm{\eta}_k=\tfrac12\Re\{G_{ik}-S_{ik}\},\nonumber\\
\bm{\xi}_i^T\bm{\eta}_k&=\tfrac12\Im\{G_{ik}+S_{ik}\}.
\label{eq:app4}
\end{align}
These relations are consistent since $\vect{G}$ is Hermitian and $\vect{S}$ symmetric, and define a real symmetric $2K\times2K$ matrix $\vect{R}_\mathrm{e}$. With the invertible $\vect{T}=\big[\begin{smallmatrix}\vect{I}_K&\vect{I}_K\\ j\vect{I}_K&-j\vect{I}_K\end{smallmatrix}\big]$, $[\widetilde{\vect{U}},\widetilde{\vect{U}}^*]=[\bm{\Xi},\bm{\Upsilon}]\vect{T}$ with $\bm{\Xi}=[\bm{\xi}_1,\dots,\bm{\xi}_K]$, $\bm{\Upsilon}=[\bm{\eta}_1,\dots,\bm{\eta}_K]$, so $\vect{R}_\mathrm{e}=\vect{T}^{-H}\vect{K}\vect{T}^{-1}\succeq\vect{0}$. Factorizing $\vect{R}_\mathrm{e}=\vect{L}^T\vect{L}$ and embedding the $\mathrm{rank}(\vect{R}_\mathrm{e})=\mathrm{rank}[\vect{V},\vect{U}^*]\leq M$ nonzero rows of $\vect{L}$ into $\mathbb{R}^M$ yields real vectors with the prescribed Gram matrix, hence $\widetilde{\vect{U}}$ satisfying~\eqref{eq:app3}; Steps 3 and 1 give $\vect{Q}$ and $\Th$.

\section{Proof of Lemma~\ref{lem:inc}}
\label{app:inc}
Since $\vect{H}=\vect{E}\bm{\Lambda}^{1/2}\vect{W}^H$, every incident field $\vect{H}\vect{w}$ lies in $\mathrm{span}(\vect{E})$. For a target $\vect{u}=\vect{E}\vect{t}$ with $\lambda_2>0$, $\vect{H}\vect{w}=\vect{E}\vect{t}$ is equivalent to $\vect{W}^H\vect{w}=\bm{\Lambda}^{-1/2}\vect{t}$. Since $\|\vect{w}\|^2\geq\|\vect{W}^H\vect{w}\|^2$, with equality if and only if $\vect{w}\in\mathrm{span}(\vect{W})$, the minimum power is $\|\bm{\Lambda}^{-1/2}\vect{t}\|^2=\vect{t}^H\bm{\Lambda}^{-1}\vect{t}$, attained by $\vect{w}=\vect{W}\bm{\Lambda}^{-1/2}\vect{t}$, and the bounds follow from $\lambda_2\leq\lambda_i\leq\lambda_1$. For the eigenvalues, the nonzero eigenvalues of $\vect{H}\vect{H}^H=\vect{B}\vect{D}_0^{1/2}\vect{A}^H\vect{A}\vect{D}_0^{1/2}\vect{B}^H=N\vect{B}\vect{D}_0\vect{B}^H$ coincide with those of the $2\times2$ matrix
\begin{equation}
N\vect{D}_0^{1/2}\vect{B}^H\vect{B}\vect{D}_0^{1/2}=NM\begin{bmatrix}\beta_{0,1}&\sqrt{\beta_{0,1}\beta_{0,2}}\,r_0^*\\ \sqrt{\beta_{0,1}\beta_{0,2}}\,r_0&\beta_{0,2}\end{bmatrix},
\end{equation}
with $r_0=A(\vect{p};\dsin^0)$ and $|r_0|=\rho_0$. Its trace is $NM(\beta_{0,1}+\beta_{0,2})$ and its determinant $N^2M^2\beta_{0,1}\beta_{0,2}(1-\rho_0^2)$, which give~\eqref{eq:lam}.

\section{Proof of Lemma~\ref{lem:phase}}
\label{app:phase}
Let $\vect{y}_k=\vect{R}^H\vect{Q}_0^H\vect{d}_k\in\mathbb{C}^2$, which has unit norm since $\vect{d}_k\in\mathrm{span}(\vect{Q}_0)$. From~\eqref{eq:frame}, $\vect{F}(\phi)^H\vect{d}_k=\bm{\Phi}^*\mathrm{diag}(1,e^{-j\phi})\vect{y}_k$. Hence
\begin{align}
\pi_k(\phi)&=\sum_{i=1}^2\frac{\big|y_{k,1}\Phi_{i1}^*+y_{k,2}\Phi_{i2}^*e^{-j\phi}\big|^2}{\lambda_i}\nonumber\\
&=c_{k,0}+2\Re\{c_{k,1}e^{-j\phi}\},
\end{align}
with $c_{k,0}=\sum_i(|y_{k,1}|^2|\Phi_{i1}|^2+|y_{k,2}|^2|\Phi_{i2}|^2)/\lambda_i$ and $c_{k,1}=\sum_iy_{k,1}^*\Phi_{i1}y_{k,2}\Phi_{i2}^*/\lambda_i$, which is the stated sinusoid. Its extreme values are costs of unit-norm vectors $\vect{F}^H\vect{d}_k$ and therefore lie in $[1/\lambda_1,1/\lambda_2]$.

\section{Proof of Proposition~\ref{prop:dn}}
\label{app:dn}
\emph{Feasibility.} Blocks have length $(M_b-1)d_0$, and consecutive offsets differ by $\Delta_0\geq M_bd_0$ or by $\Delta_1-\Delta_0\geq M_bd_0$, so the spacing is at least $d_0$; the last element is at $D_\mathrm{dn}\leq D$.

\emph{Nulls.} Let $A_b(s)$ be the array factor of one block. The placement is the sum of four shifted blocks, so
\begin{equation}
A(\vect{p};s)=\tfrac14A_b(s)\big(1+e^{j\kappa s\Delta_0}\big)\big(1+e^{j\kappa s\Delta_1}\big).
\end{equation}
By construction, $\kappa\dsin\Delta_1=2\pi\dsin\Delta_1/\lambda$ is an odd multiple of $\pi$, so the third factor vanishes at $s=\dsin$, and likewise the second at $s=\dsin^0$. Hence $\rho=\rho_0=0$. By~\eqref{eq:lam} with equal gains, $\lambda_1=\lambda_2=\beta_0NM$, and~\eqref{eq:twoiso} gives~\eqref{eq:dn}.

\emph{Converse.} The phases $\kappa\dsin p_m$, $p_m\in[0,D]$, lie in an arc of width $\kappa|\dsin|D<\pi$; projecting the unit phasors onto its bisector gives $M\rho\geq M\cos(\kappa|\dsin|D/2)=M\cos(\pi|\dsin|D/\lambda)$.

\section{Proof of Proposition~\ref{prop:small}}
\label{app:small}
Let $x_m=\kappa\dsin(p_m-\bar p)$, so that $\rho=|\frac1M\sum_me^{jx_m}|$ and $\sum_mx_m=0$. Expanding, $\frac1M\sum_me^{jx_m}=1-\frac{1}{2M}\sum_mx_m^2+O(x^3)$ with $|x_m|\leq\kappa|\dsin|D$, and the cubic terms cancel in $\rho^2=1-\kappa^2\dsin^2\Var(\vect{p})+O(x^4)$, which proves~\eqref{eq:small}. $\Var(\vect{p})$ is a convex quadratic, so its maximum over the polytope $\mathcal{X}$ is attained at a vertex, where all gaps but one equal $d_0$: two compact clusters of $k$ and $M-k$ elements at the ends of the track, with centroid distance $c_k=D-(M-2)d_0/2$. Their variance is $f(k)=\frac{k(M-k)}{M^2}c_k^2+\frac{d_0^2}{12M}\big(k^3+(M-k)^3-M\big)$, with $f''(k)=-2c_k^2/M^2+d_0^2/2<0$ since $c_k>Md_0/2$ when $D>(M-1)d_0$; by symmetry, $f$ is maximized at $k=M/2$, which gives~\eqref{eq:varmax}. The compact array has the variance $d_0^2(M^2-1)/12$ of $M$ equispaced points.

\section{Proof of Proposition~\ref{prop:leak}}
\label{app:leak}
Expanding $\vect{a}(\hat\theta+\nu)=\vect{a}+\nu\dot{\vect{a}}+\frac{\nu^2}2\ddot{\vect{a}}+O(\nu^3)$ at $\hat\theta$ gives $\vect{a}(\hat\theta+\nu)^H\vect{d}_1=\vect{a}^H\vect{d}_1+\nu\dot{\vect{a}}^H\vect{d}_1+\frac{\nu^2}2\ddot{\vect{a}}^H\vect{d}_1+O(\nu^3)$. (a) For the nominal direction, $\vect{a}^H\vect{d}_1=0$, so $\zeta=\nu^2|\dot{\vect{a}}^H\vect{d}_1|^2+O(\nu^3)$. Since $\vect{d}_1\perp\vect{a}$ and $\|\vect{d}_1\|=1$, $|\dot{\vect{a}}^H\vect{d}_1|=|\dot{\vect{a}}_\perp^H\vect{d}_1|\leq\|\dot{\vect{a}}_\perp\|$, and
\begin{equation}
\|\dot{\vect{a}}_\perp\|^2=\|\dot{\vect{a}}\|^2-\frac{|\vect{a}^H\dot{\vect{a}}|^2}{M}=\kappa^2\cos^2\hat\theta\Big(\sum_mp_m^2-\frac{(\sum_mp_m)^2}{M}\Big),
\end{equation}
which equals $\kappa^2\cos^2\hat\theta\,M\Var(\vect{p})$, because $|[\vect{a}]_m|=1$, $[\dot{\vect{a}}]_m=j\kappa\cos\hat\theta\,p_m[\vect{a}]_m$, and $\vect{a}^H\dot{\vect{a}}=j\kappa\cos\hat\theta\sum_mp_m$. (b) If $\vect{g}_\mathrm{B}\notin\mathrm{span}(\vect{Z})$, for the derivative-null direction, $\vect{a}^H\vect{d}_1^\mathrm{rob}=\dot{\vect{a}}^H\vect{d}_1^\mathrm{rob}=0$, so the first nonzero term is quadratic in $e$, which gives the stated expansion. Finally, $\vect{P}_{\vect{Z}}^\perp=\vect{P}_{\vect{a}}^\perp-\dot{\vect{a}}_\perp\dot{\vect{a}}_\perp^H/\|\dot{\vect{a}}_\perp\|^2$, because $\{\vect{a}/\sqrt{M},\dot{\vect{a}}_\perp/\|\dot{\vect{a}}_\perp\|\}$ is an orthonormal basis of $\mathrm{span}(\vect{Z})$, and hence $|\vect{g}_\mathrm{B}^H\vect{d}_1^\mathrm{rob}|^2=\|\vect{P}_{\vect{Z}}^\perp\vect{g}_\mathrm{B}\|^2=\|\vect{g}_\mathrm{B}\|^2(1-\rho^2-\dot\rho^2)$ with $\rho$ evaluated for the LoS target channel.

{\footnotesize
\bibliographystyle{IEEEtran}
\def\baselinestretch{0.85}
\bibliography{refs_tifs}}

@inproceedings{sheemar2023full_ris,
  author    = {C. K. Sheemar and G. C. Alexandropoulos and
               D. Slock and J. Querol and S. Chatzinotas},
  title     = {Full-Duplex-Enabled Joint Communications and
               Sensing with Reconfigurable Intelligent Surfaces},
  booktitle = {Proc. 31st Eur. Signal Process. Conf. (EUSIPCO)},
  pages     = {1509--1513},
  year      = {2023}
}

@article{sheemar2025holographic,
  title={Holographic joint communications and sensing with cramer-rao bounds},
  author={Sheemar, Chandan Kumar and Khan, Wali Ullah and Alexandropoulos, George C and Querol, Jorge and Chatzinotas, Symeon},
  journal={IEEE Journal on Selected Areas in Communications},
  volume={44},
  pages={498--512},
  year={2025},
  publisher={IEEE}
}

@article{li2025ris,
  title={RIS-based physical layer security for integrated sensing and communication: A comprehensive survey},
  author={Li, Yongxiao and Khan, Feroz and Ahmed, Manzoor and Soofi, Aized Amin and Khan, Wali Ullah and Sheemar, Chandan Kumar and Asif, Muhammad and Han, Zhu},
  journal={IEEE Internet of Things Journal},
  volume={12},
  number={16},
  pages={32444--32468},
  year={2025},
  publisher={IEEE}
}

@article{khan2026survey,
  title={Survey on beyond diagonal RIS enabled 6G wireless networks: Fundamentals, recent advances, and challenges},
  author={Khan, Wali Ullah and Sheemar, Chandan Kumar and Ahmed, Manzoor and Di Renzo, Marco and Shah, Syed Tariq and Mahmood, Asad and Lagunas, Eva and Querol, Jorge and Chatzinotas, Symeon and Han, Zhu},
  journal={IEEE Open Journal of the Communications Society},
  year={2026},
  publisher={IEEE}
}

@article{iacovelli2026mutual,
  title={Mutual-Coupling-Aware Movable and Fluid Antennas on Holographic Surfaces: A Wavenumber-Domain Circuit-Field Unification},
  author={Iacovelli, Giovanni and Sheemar, Chandan Kumar and Chatzinotas, Symeon},
  journal={arXiv preprint arXiv:2609.04968},
  year={2026}
}

@article{Su2021,
  author  = {N. Su and F. Liu and C. Masouros},
  title   = {Secure radar-communication systems with malicious targets: Integrating radar, communications and jamming functionalities},
  journal = {IEEE Trans. Wireless Commun.},
  volume  = {20}, number = {1}, pages = {83--95}, year = {2021},
  doi     = {10.1109/TWC.2020.3023164}
}

@article{Cui2019,
  author  = {M. Cui and G. Zhang and R. Zhang},
  title   = {Secure wireless communication via intelligent reflecting surface},
  journal = {IEEE Wireless Commun. Lett.},
  volume  = {8}, number = {5}, pages = {1410--1414}, year = {2019},
  doi     = {10.1109/LWC.2019.2919685}
}

@article{Shen2022,
  author  = {S. Shen and B. Clerckx and R. Murch},
  title   = {Modeling and architecture design of reconfigurable intelligent surfaces using scattering parameter network analysis},
  journal = {IEEE Trans. Wireless Commun.},
  volume  = {21}, number = {2}, pages = {1229--1243}, year = {2022},
  doi     = {10.1109/TWC.2021.3103256}
}

@article{Li2023modes,
  author  = {H. Li and S. Shen and B. Clerckx},
  title   = {Beyond diagonal reconfigurable intelligent surfaces: From transmitting and reflecting modes to single-, group-, and fully-connected architectures},
  journal = {IEEE Trans. Wireless Commun.},
  volume  = {22}, number = {4}, pages = {2311--2324}, year = {2023},
  doi     = {10.1109/TWC.2022.3210706}
}

@article{Nerini2024closed,
  author  = {M. Nerini and S. Shen and B. Clerckx},
  title   = {Closed-form global optimization of beyond diagonal reconfigurable intelligent surfaces},
  journal = {IEEE Trans. Wireless Commun.},
  volume  = {23}, number = {2}, pages = {1037--1051}, year = {2024},
  doi     = {10.1109/TWC.2023.3285262}
}

@article{Zhang2025,
  author  = {X. Zhang and L. Liu and S. Zhang and W. Zhu and H. Zhang},
  title   = {Optimizing rate-{CRB} performance for beyond diagonal reconfigurable intelligent surface enabled {ISAC}},
  journal = {IEEE Commun. Lett.},
  volume  = {29}, number = {11}, pages = {2506--2510}, year = {2025},
  doi     = {10.1109/LCOMM.2025.3599597}
}

@article{Zhao2026,
  author  = {J. Zhao and Q. Huang and K. Cai and Q. Zhou and X. Mu and Y. Liu},
  title   = {Movable-element {RIS}-aided wireless communications: An element-wise position optimization approach},
  journal = {IEEE Commun. Lett.},
  volume  = {30}, pages = {967--971}, year = {2026},
  doi     = {10.1109/LCOMM.2026.3655084}
}

@article{Xu2026,
  author  = {S. Xu and M. Nerini and B. Clerckx},
  title   = {Movable beyond-diagonal reconfigurable intelligent surfaces: Moving, interconnecting, or both?},
  journal = {IEEE Wireless Commun. Lett.},
  volume  = {15}, pages = {2144--2148}, year = {2026},
  doi     = {10.1109/LWC.2026.3672039}
}

@article{Wyner1975,
  author  = {A. D. Wyner},
  title   = {The wire-tap channel},
  journal = {Bell Syst. Tech. J.},
  volume  = {54}, number = {8}, pages = {1355--1387}, year = {1975},
  doi     = {10.1002/j.1538-7305.1975.tb02040.x}
}

@article{Csiszar1978,
  author  = {I. Csisz{\'a}r and J. K{\"o}rner},
  title   = {Broadcast channels with confidential messages},
  journal = {IEEE Trans. Inf. Theory},
  volume  = {24}, number = {3}, pages = {339--348}, year = {1978},
  doi     = {10.1109/TIT.1978.1055892}
}

@article{Goel2008,
  author  = {S. Goel and R. Negi},
  title   = {Guaranteeing secrecy using artificial noise},
  journal = {IEEE Trans. Wireless Commun.},
  volume  = {7}, number = {6}, pages = {2180--2189}, year = {2008},
  doi     = {10.1109/TWC.2008.060848}
}

@article{Khisti2010,
  author  = {A. Khisti and G. W. Wornell},
  title   = {Secure transmission with multiple antennas {I}: The {MISOME} wiretap channel},
  journal = {IEEE Trans. Inf. Theory},
  volume  = {56}, number = {7}, pages = {3088--3104}, year = {2010},
  doi     = {10.1109/TIT.2010.2048445}
}

@article{Guan2020,
  author  = {X. Guan and Q. Wu and R. Zhang},
  title   = {Intelligent reflecting surface assisted secrecy communication: Is artificial noise helpful or not?},
  journal = {IEEE Wireless Commun. Lett.},
  volume  = {9}, number = {6}, pages = {778--782}, year = {2020},
  doi     = {10.1109/LWC.2020.2969629}
}

@article{Yu2020,
  author  = {X. Yu and D. Xu and Y. Sun and D. W. K. Ng and R. Schober},
  title   = {Robust and secure wireless communications via intelligent reflecting surfaces},
  journal = {IEEE J. Sel. Areas Commun.},
  volume  = {38}, number = {11}, pages = {2637--2652}, year = {2020},
  doi     = {10.1109/JSAC.2020.3007043}
}

@article{Su2022,
  author  = {N. Su and F. Liu and Z. Wei and Y.-F. Liu and C. Masouros},
  title   = {Secure dual-functional radar-communication transmission: Exploiting interference for resilience against target eavesdropping},
  journal = {IEEE Trans. Wireless Commun.},
  volume  = {21}, number = {9}, pages = {7238--7252}, year = {2022},
  doi     = {10.1109/TWC.2022.3156893}
}

@article{Su2024,
  author  = {N. Su and F. Liu and C. Masouros},
  title   = {Sensing-assisted eavesdropper estimation: An {ISAC} breakthrough in physical layer security},
  journal = {IEEE Trans. Wireless Commun.},
  volume  = {23}, number = {4}, pages = {3162--3174}, year = {2024},
  doi     = {10.1109/TWC.2023.3306029}
}

@article{Wei2022,
  author  = {Z. Wei and F. Liu and C. Masouros and N. Su and A. P. Petropulu},
  title   = {Toward multi-functional {6G} wireless networks: Integrating sensing, communication, and security},
  journal = {IEEE Commun. Mag.},
  volume  = {60}, number = {4}, pages = {65--71}, year = {2022},
  doi     = {10.1109/MCOM.002.2100972}
}

@article{LiuX2020,
  author  = {X. Liu and T. Huang and N. Shlezinger and Y. Liu and J. Zhou and Y. C. Eldar},
  title   = {Joint transmit beamforming for multiuser {MIMO} communications and {MIMO} radar},
  journal = {IEEE Trans. Signal Process.},
  volume  = {68}, pages = {3929--3944}, year = {2020},
  doi     = {10.1109/TSP.2020.3004739}
}

@article{Hua2023,
  author  = {H. Hua and J. Xu and T. X. Han},
  title   = {Optimal transmit beamforming for integrated sensing and communication},
  journal = {IEEE Trans. Veh. Technol.},
  volume  = {72}, number = {8}, pages = {10588--10603}, year = {2023},
  doi     = {10.1109/TVT.2023.3262513}
}

@article{Ren2023robust,
  author  = {Z. Ren and L. Qiu and J. Xu and D. W. K. Ng},
  title   = {Robust transmit beamforming for secure integrated sensing and communication},
  journal = {IEEE Trans. Commun.},
  volume  = {71}, number = {9}, pages = {5549--5564}, year = {2023},
  doi     = {10.1109/TCOMM.2023.3286461}
}

@article{Chu2023,
  author  = {J. Chu and R. Liu and M. Li and Y. Liu and Q. Liu},
  title   = {Joint secure transmit beamforming designs for integrated sensing and communication systems},
  journal = {IEEE Trans. Veh. Technol.},
  volume  = {72}, number = {4}, pages = {4778--4791}, year = {2023},
  doi     = {10.1109/TVT.2022.3225952}
}

@article{Hua2024secure,
  author  = {M. Hua and Q. Wu and W. Chen and O. A. Dobre and A. L. Swindlehurst},
  title   = {Secure intelligent reflecting surface-aided integrated sensing and communication},
  journal = {IEEE Trans. Wireless Commun.},
  volume  = {23}, number = {1}, pages = {575--591}, year = {2024},
  doi     = {10.1109/TWC.2023.3280179}
}

@article{Wei2024STAR,
  author  = {W. Wei and X. Pang and C. Xing and N. Zhao and D. Niyato},
  title   = {{STAR-RIS} aided secure {NOMA} integrated sensing and communication},
  journal = {IEEE Trans. Wireless Commun.},
  volume  = {23}, number = {9}, pages = {10712--10725}, year = {2024},
  doi     = {10.1109/TWC.2024.3374728}
}

@article{ZhaoH2026,
  author  = {H. Zhao and M. Wu and D. Wang and M. Guizani and V. C. M. Leung},
  title   = {Sensing-assisted secure beamforming for {RIS}-enabled {ISAC} with leakage suppression},
  journal = {IEEE Trans. Wireless Commun.},
  volume  = {25}, pages = {17755--17769}, year = {2026},
  doi     = {10.1109/TWC.2026.3695639}
}

@article{Nerini2024graph,
  author  = {M. Nerini and S. Shen and H. Li and B. Clerckx},
  title   = {Beyond diagonal reconfigurable intelligent surfaces utilizing graph theory: Modeling, architecture design, and optimization},
  journal = {IEEE Trans. Wireless Commun.},
  volume  = {23}, number = {8}, pages = {9972--9985}, year = {2024},
  doi     = {10.1109/TWC.2024.3367631}
}

@article{Xiong2025,
  author  = {W. Xiong and J. Lin and C. Pan and Y. Zeng and Q. Li},
  title   = {Enhancing physical layer security in {MIMO} systems assisted by beyond-diagonal reconfigurable intelligent surfaces},
  journal = {IEEE Trans. Commun.},
  volume  = {73}, number = {11}, pages = {10824--10839}, year = {2025},
  doi     = {10.1109/TCOMM.2025.3592603}
}

@article{ZhangX2025PLS,
  author  = {X. Zhang and X. Ma and H. Zhang and Y. Ren and D. Cao and V. C. M. Leung},
  title   = {Physical layer security enhancement for beyond diagonal {RIS}-aided {MIMO} communications},
  journal = {IEEE Commun. Lett.},
  volume  = {29}, number = {8}, pages = {1919--1923}, year = {2025},
  doi     = {10.1109/LCOMM.2025.3579860}
}

@article{Liu2024BDISAC,
  author  = {Z. Liu and Y. Liu and S. Shen and Q. Wu and Q. Shi},
  title   = {Enhancing {ISAC} network throughput using beyond diagonal {RIS}},
  journal = {IEEE Wireless Commun. Lett.},
  volume  = {13}, number = {6}, pages = {1670--1674}, year = {2024},
  doi     = {10.1109/LWC.2024.3386155}
}

@article{ZhangX2025CRB,
  author  = {X. Zhang and L. Liu and S. Zhang and H. Zhang},
  title   = {Beyond diagonal reconfigurable intelligent surface enabled sensing: {Cram\'er-Rao} bound optimization},
  journal = {IEEE Wireless Commun. Lett.},
  volume  = {14}, number = {12}, pages = {3852--3856}, year = {2025},
  doi     = {10.1109/LWC.2025.3599590}
}

@misc{Illi2026,
  author       = {E. Illi and M. Qaraqe},
  title        = {Beyond-diagonal {RIS} for enhanced secrecy and sensing gains in secure {ISAC} networks: An optimization framework},
  howpublished = {arXiv preprint arXiv:2604.04480},
  year         = {2026}
}

@article{Ma2024MIMO,
  author  = {W. Ma and L. Zhu and R. Zhang},
  title   = {{MIMO} capacity characterization for movable antenna systems},
  journal = {IEEE Trans. Wireless Commun.},
  volume  = {23}, number = {4}, pages = {3392--3407}, year = {2024},
  doi     = {10.1109/TWC.2023.3307696}
}

@article{Zhu2024model,
  author  = {L. Zhu and W. Ma and R. Zhang},
  title   = {Modeling and performance analysis for movable antenna enabled wireless communications},
  journal = {IEEE Trans. Wireless Commun.},
  volume  = {23}, number = {6}, pages = {6234--6250}, year = {2024},
  doi     = {10.1109/TWC.2023.3330887}
}

@article{Hu2024,
  author  = {G. Hu and Q. Wu and K. Xu and J. Si and N. Al-Dhahir},
  title   = {Secure wireless communication via movable-antenna array},
  journal = {IEEE Signal Process. Lett.},
  volume  = {31}, pages = {516--520}, year = {2024},
  doi     = {10.1109/LSP.2024.3359894}
}

@article{Lyu2025MAISAC,
  author  = {W. Lyu and S. Yang and Y. Xiu and Z. Zhang and C. Assi and C. Yuen},
  title   = {Movable antenna enabled integrated sensing and communication},
  journal = {IEEE Trans. Wireless Commun.},
  volume  = {24}, number = {4}, pages = {2862--2875}, year = {2025},
  doi     = {10.1109/TWC.2025.3525631}
}

@article{Ma2025secure,
  author  = {Y. Ma and K. Liu and Y. Liu and L. Zhu and Z. Xiao},
  title   = {Movable-antenna aided secure transmission for {RIS-ISAC} systems},
  journal = {IEEE Trans. Wireless Commun.},
  volume  = {24}, number = {12}, pages = {10019--10035}, year = {2025},
  doi     = {10.1109/TWC.2025.3577040}
}

@article{Zhu2025FRIS,
  author  = {X. Zhu and K.-K. Wong and B. Tang and W. Chen and C.-B. Chae},
  title   = {Fluid reconfigurable intelligent surface ({FRIS}) enabling secure wireless communications},
  journal = {IEEE Wireless Commun. Lett.},
  volume  = {15}, pages = {2408--2412}, year = {2026},
  doi     = {10.1109/LWC.2026.3676856}
}

@article{Zhuang2026,
  author  = {L. Zhuang and X. Xie and F. Fang and A. Attaran and Z. Zhang},
  title   = {Robust and secure transmission for movable-{RIS}-assisted {ISAC} with imperfect sense estimation},
  journal = {IEEE Trans. Wireless Commun.},
  volume  = {25}, pages = {17071--17085}, year = {2026},
  doi     = {10.1109/TWC.2026.3693068}
}

@book{HornJohnson2013,
  author    = {R. A. Horn and C. R. Johnson},
  title     = {Matrix Analysis},
  edition   = {2nd},
  publisher = {Cambridge Univ. Press},
  address   = {Cambridge, U.K.},
  year      = {2013}
}

\end{document}